\documentclass{jpp}
\usepackage{graphicx}
\usepackage[T1]{fontenc}
\usepackage{amsmath}
\usepackage{bm}
\usepackage{natbib}
\usepackage{float}
\usepackage[colorlinks=true,
  linkcolor=black, citecolor=blue, urlcolor=blue]{hyperref}   
\usepackage[dvipsnames,svgnames,x11names,hyperref]{xcolor}  
\usepackage[capitalize]{cleveref}

\usepackage{setspace}
\crefformat{subsection}{\S#1}

\shorttitle{Gyrokinetic moment-based simulations of the
Dimits shift}
\shortauthor{A C.D. Hoffmann et al.}

\title{Gyrokinetic moment-based simulations of the Dimits shift}
\author{A.C.D. Hoffmann \aff{1}\corresp{\email{antoine.hoffmann@epfl.ch}}, B.J. Frei \aff{1}, P. Ricci \aff{1}}
\affiliation{\aff{1} Ecole Polytechnique F\'ed\'erale de Lausanne (EPFL), Swiss
Plasma Center, CH-1015 Lausanne, Switzerland
}
 \newcommand{\squareparenthesis}[1]{\left[#1\right]} 
 \newcommand{\roundparenthesis}[1]{\left(#1\right)} 
 \newcommand{\curlyparenthesis}[1]{\left\{#1\right\}} 
 \newcommand{\refeq}[1]{Eq. \eqref{#1}}

 \newcommand{\kperp}{k_\perp}

 \newcommand{\vpar}{v_\parallel}
 \newcommand{\Nvpar}{N_{v_\parallel}}

 \newcommand{\vtha}{v_{tha}}

 \newcommand{\phibar}{\bar \phi}

 \newcommand{\dz}{\mathrm dz}
 \newcommand{\ddt}{\partial_t}

 \newcommand{\ddx}{\partial_x}
 \newcommand{\ddy}{\partial_y}
 \newcommand{\ddz}{\partial_z}
 \newcommand{\ddvpar}{\partial_{v\parallel}}
 \newcommand{\ddspar}{\partial_{s\parallel}}

 \newcommand{\grad}{\nabla}
 \newcommand{\gradpar}{\nabla_\parallel}
 \newcommand{\bvec}{\bm{b}}

\newcommand{\kernel}{\mathcal{K}}

\newcommand{\kT}{\kappa_{T}}

\newcommand{\modifi}[1]{{#1}}
\newcommand{\modifii}[1]{{#1}}
\newcommand{\modifiii}[1]{#1}
\newcommand{\modifiv}[1]{#1}
\newcommand{\reviewi}[1]{#1}
\newcommand{\reviewii}[1]{#1}
\newcommand{\reviewiii}[1]{#1}

\usepackage{natbib}
\usepackage{graphicx}
\usepackage{bm}
\usepackage{amsmath}
\numberwithin{equation}{section}
\usepackage{amssymb}
\usepackage{cancel}
\usepackage{float}
\usepackage{algorithmic}
\usepackage{algorithm}
\usepackage{multicol}

\begin{document}

\maketitle

\begin{abstract}
\reviewi{We present a convergence study of the gyromoment (GM) approach, which is based on projecting the gyrokinetic distribution function onto a Hermite-Laguerre polynomial basis, focused on the cyclone base case (CBC) \citep{Lin1999EffectsTransport} and Dimits shift \citep{Dimits2000ComparisonsSimulations} as benchmarks.}
We report that the GM approach converges more rapidly in capturing the nonlinear dynamics of the CBC than the continuum GENE code \citep{Jenko2000} when comparing the number of points representing the velocity space. 
Increasing the velocity dissipation improves the convergence properties of the GM approach, albeit yielding a slightly larger saturated heat flux. 
By varying the temperature equilibrium gradient, we show that GM approach successfully reproduces the Dimits shift \citep{Dimits2000ComparisonsSimulations} and effectively captures its width, which is in contrast to the gyrofluid framework. 
In the collisional regime, the convergence properties of the GM approach improve and a good agreement with previous global PIC results on transport is obtained \citep{Lin1999EffectsTransport}. 
Finally, we report that the choice of collision model has a minimal impact both on the ITG growth rate and on the nonlinear saturated heat flux, at tokamak-relevant collisionality.
\end{abstract}

\section{Introduction}
To reduce the computational cost of gyrokinetic (GK) simulations, gyrofluid (GF) models evolve a limited number of \reviewi{moments of the distribution function, according to conservation laws derived from the GK Boltzmann equation}  \citep{Grant1967Fourier-HermiteLimit,Madsen2013Full-FModel,Held2020Pade-basedModels}.
However, a comparative study with a set of GK codes presented by \cite{Dimits2000ComparisonsSimulations} reveals that the use of GF models can lead to erroneous results.
Focusing on the cyclone base case (CBC), a simulation scenario with the parameters of a DIII-D H-mode discharge \citep{Greenfield1997EnhancedDIII-D}, and using particle-in-cell (PIC) and continuum GK codes, \cite{Dimits2000ComparisonsSimulations} identify a shift between the minimum value of the background temperature gradient that yields a finite saturated nonlinear heat flux, and the linear ion temperature gradient (ITG) stability threshold.
In contrast, the GF model of \cite{Beer1995} does not predict the Dimits shift, thus questioning of the validity of GF models for tokamak core simulations.
In fact, despite a correct prediction of the ITG linear growth rate at conventional CBC parameters, the GF model does not present a strongly reduced transport in the ITG marginal stability region, overestimating drastically the transport in comparison to GK codes.
\par
\modifi{In this study, we show that the limitations of the GF models revealed in \cite{Dimits2000ComparisonsSimulations} \modifii{can be overcome with a GK moment-based approach \citep{Frei2020}}.
This approach extends the drift-kinetic (DK) model presented in \cite{Jorge2017ACollisionality} and is based on the projection of the velocity space dependence of the distribution function on a Hermite-Laguerre polynomial basis at an arbitrary order.}
This yields an infinite set of fluid-like equations for the basis coefficients, referred to as gyromoment (GMs), \modifii{that extend the GF models to an arbitrary number of moments}.
As the number of GK Hermite-Laguerre moments increases, the evolution of the distribution function converges to the one provided by the full-F GK Boltzmann equation.
In addition, \cite{Jorge2019b} project advanced GK collision operators on the same basis, including the nonlinear GK Fokker-Planck collision operator \citep{Rosenbluth1957}, thus providing a GK model that includes \modifii{an advanced description of collisional effects} and combines the accuracy of the GK model with the efficiency of a fluid approach, in particular when a high collisionality regime is considered.
\cite{Frei2020} introduce the linear $\delta f$ flux-tube limit of the GK moment approach, where equilibrium and fluctuating quantities are separated as in the simulation of the CBC and Dimits shift. 
The $\delta f$ flux-tube GM model can be considered an extension of \modifii{$\delta f$ flux-tube GF models}, as the ones introduced by \cite{Brizard1992NonlinearPlasmas}, \cite{Hammett1992FluidDynamics}, \cite{Beer1995}, \cite{Snyder2001AMicroturbulence} and \cite{scott2005}, thanks to the use of an arbitrary number of GMs to describe the evolution of the perturbations of a Maxwellian equilibrium distribution.
Within the $\delta f$ framework, \cite{Frei2022LocalMode} also project and use several GK collision operators, in particular the Dougherty model \citep{Dougherty1964}, the Sugama model \citep{Sugama2006} and the Landau form of the full Coulomb collision model \citep{Rosenbluth1957Fokker-planckForce,Hazeltine2003PlasmaConfinement}.
\par
Using the linear $\delta f$ GM model, \cite{Frei2023Moment-basedModel} validate the GM approach with investigations of toroidal ITG modes, trapped electron modes (TEM), microtearing modes (MTM) and kinetic ballooning modes (KBM) in the $s-\alpha$ flux-tube geometry. 
\cite{Hoffmann2023GyrokineticOperators} present the first nonlinear simulations based on the $\delta f$ GM model and use them to investigate the evolution of turbulence in a Z-pinch configuration, where the dynamics is dominated by zonal flows (ZF), considering both the collisionless and the collisional cases.
The number of GMs needed for convergence in the collisionless case is shown to be less than the number of grid points necessary for convergence in the state-of-the-art continuum GK code GENE \citep{Jenko2000}, depending on the background gradient strength.
Moreover, the results reveal that the choice of the collision model significantly impacts the level of saturated transport, even at low collision frequency. 
\par
A similar nonlinear  $\delta f$ GM model is presented by \cite{Mandell2018} and \cite{Mandell2022GX:Design} and implemented in GX, a GPU native code that solves the nonlinear $\delta f$ GM hierarchy of equations in record times.
GX is benchmarked against GS2 \citep{kotschenreuther1995}, CGYRO \citep{Candy2003AnSolver}, and Stella \citep{Barnes2019Stella:Configurations} in linear and nonlinear cases, for tokamak Miller \citep{Miller1998NoncircularModel} and stellarator magnetic equilibria, while considering both adiabatic and kinetic electrons.
In contrast to GF models, \cite{Mandell2022GX:Design} show that the GM approach can retrieve the Dimits shift.
\par
Extending the work in \cite{Mandell2022GX:Design}, the present study aims to investigate the convergence properties of the GM method in the simulation of the CBC and the Dimits shift \citep{Dimits2000ComparisonsSimulations} using the GYACOMO code \citep{gyacomo}.
In addition, the impact of the collision model on the CBC is studied by comparing different linear GK collision models, thus extending the work of \cite{Lin1999EffectsTransport}, which \reviewi{studies} heat transport close to the Dimits threshold using a GK particle-in-cell code and a pitch-angle collision operator at a physically relevant collision frequency.
Finally, the effect of collisions on the Dimits shift is investigated.
\par
%
Our study confirms that, similarly to GX, the $\delta f$ GM approach converges to the correct ITG linear stability threshold and nonlinear transport value, which sets it apart from GF models.
This is assessed by comparing the results of the GM approach with the \cite{Dimits2000ComparisonsSimulations} results as well as results from the GENE code \cite{Jenko2000}. 
We also conduct a detailed convergence study with respect to the velocity space resolution.
We analyze the impact of the background temperature gradient levels, the number of GMs, and the level of numerical velocity dissipation on the convergence properties.
Finally, we focus on a specific temperature gradient value corresponding to the Dimits regime and investigate the impact of collisions on the level of transport \citep{Lin1999EffectsTransport}\modifii{, as well as the impact of collisions on the Dimits shift.} 
We compare the results obtained using three different GK collision operators: Dougherty \citep{Dougherty1964}, Sugama \citep{Sugama2009LinearizedEquations}, and Landau \citep{Rosenbluth1957Fokker-planckForce}.
\par
The paper is organized as follows. 
Section \ref{sec:gk_model} introduces the Hermite-Laguerre GM approach within the context of the CBC. 
In Sec. \ref{sec:CBC_benchmark}, we present the benchmarks of the GM results for the ITG threshold and turbulence in the collisionless limit. 
The convergence study of the Dimits shift is presented in Sec. \ref{sec:dimits_shift}. 
In Sec. \ref{sec:coll_dimits_shift}, we investigate the effect of collisions. 
The conclusions follow in Sec. \ref{sec:conclusion}.

\section{Gyrokinetic gyromoment approach in a flux-tube configuration}
\label{sec:gk_model}
In the present section, we introduce the GK $\delta f$ model in a field aligned coordinate system.
Then, we project the GK Boltzmann equation on a Hermite-Laguerre polynomial basis, thus obtaining a nonlinear gyromoment model, which is an extension of the linear model presented in \cite{Frei2023Moment-basedModel}.
Finally, the numerical implementation of the GM model is discussed.

\subsection{Nonlinear gyrokinetic model}
Within the GK approach \citep{Catto1978LinearizedGyro-kinetics,Frieman1982NonlinearEquilibria,Hazeltine2003PlasmaConfinement}, we consider an adiabatic electron model and evolve the ion gyroaveraged distribution function $\langle F_i\rangle (\bm R, \vpar, \mu, t)$, which expresses the probability density of finding an ion with guiding center $\bm R$, velocity parallel to the magnetic field $\vpar$, and magnetic moment $\mu$, at a time $t$.
Using the $\delta f$ approach, we decompose $\langle F_i\rangle$ as the sum of a time-independent background Maxwellian component and a perturbation $g$, i.e.
\begin{equation}
    \langle F_i\rangle(\bm R, \vpar, \mu, t) = F_0(\bm x, \vpar, \mu) + g(\bm R, \vpar, \mu, t),
\end{equation}
where the Maxwellian distribution is defined as 
\begin{equation}
F_0(\bm x, \vpar, \mu) =\frac{N}{(\pi^{1/2}v_{thi})^{3}} \exp \roundparenthesis{- \frac{m v_\parallel^2}{2T_i}-\frac{\mu B}{T_i}},
\end{equation}
with $B(\bm x)=||\bm B||$ the norm of the equilibrium magnetic field, $N=N(\bm x)$ the equilibrium density, $T_i=T_i(\bm x)$ the equilibrium temperature, $m_i$ the ion mass and $v_{thi}^2 = 2T_i/m_i$ its thermal velocity.
The background quantities depend on the particle position $\bm x$, which is related to the gyrocenter position as $\bm x = \bm R + \rho_i \bm e_\perp$, where $\rho_i$ is the ion Larmor radius and 
$\bm e_\perp=[\bm R - (\bm R \cdot \bm b)\bvec] /||\bm R - (\bm R \cdot \bm b)\bvec||$, with $\bvec=\bm B/B$.
\\
We assume small fluctuations of the distribution function, $g_i/F_0\sim \Delta \ll 1$, where the scaling parameter $\Delta$ measures the perturbation amplitude relative to the background \citep{Hazeltine2003PlasmaConfinement}.
We neglect electromagnetic fluctuations and assume that the background electrostatic potential vanishes, therefore denoting with $\phi(\bm x,t)$ the perturbed electrostatic potential, such that $e\phi/T_e\sim\Delta$ with $T_e$ the electron equilibrium temperature.
We also neglect MHD equilibrium pressure effects, namely $(4\pi \grad P/B^2)/(\bvec\times\grad B/B)\ll \Delta$, with $P$ the total pressure.
These assumptions yields the nonlinear $\delta f$ GK Boltzmann equation
\begin{align}
    \ddt g - \frac{q}{m}\gradpar\phibar\ddvpar g + \frac{1}{B}(\bvec\times\grad\phibar)\cdot\grad g 
    +\frac{m_i}{q B} \roundparenthesis{\vpar^2 + \frac{\mu B}{m}}(\bvec\times\grad \ln B)\cdot \grad h
    &\nonumber\\
    +\vpar\gradpar h - \frac{\mu B}{m_i}\gradpar \ln B\ddvpar h
    +\frac{1}{B}\bvec\times\roundparenthesis{\frac{\grad N}{N}
    +\squareparenthesis{\frac{m_i  v_\parallel^2}{2T_i}+\frac{\mu B}{T_i}-\frac{3}{2}}\frac{\grad T_i}{T_i}}\cdot\grad\phibar &= C_{ii},
    \label{eq:gyboeq}
\end{align}
where we introduce the parallel gradient operator $\gradpar = \bvec\cdot\grad$ and the non-adiabatic part of the ion distribution function perturbation $h=g+ F_0 q_i\phi/T_i $, being $q_i$ the ion charge.
In \refeq{eq:gyboeq}, $\phibar (\bm R,t)$ denotes the gyroaveraged electrostatic potential and $C_{ii} = C_{ii}(\bm R, \vpar, \mu)$ represents the ion-ion collision term.
\subsection{Field aligned coordinate system and magnetic geometry}
We use the field-aligned coordinates $\{\psi,\alpha,\chi\}$ with $\psi$ the flux surface label, $\chi$ the ballooning angle, and $\alpha$ the binormal angle defined as $\alpha = q(\psi)\chi - \varphi_{tor}$, where $q(\psi)$ is the safety factor and $\varphi_{tor}$ the toroidal angle \citep{Hazeltine2003PlasmaConfinement}.
Within these coordinates, the magnetic field can be expressed in Clebsch form, $\bm B = \nabla\psi\times\nabla\alpha$.
Considering the local limit, we evaluate the equilibrium quantities at the flux tube center position, $r_0$, defining $B_0=B(r_0)$ and $q_0 = q(\psi(r_0))$.
We then normalize the coordinates \citep{Lapillonne2009ClarificationsTurbulence,Frei2023Moment-basedModel}
\begin{equation}
    x=\frac{q_0}{r_0 B_0}[\psi-\psi(r_0)],\quad y=\frac{R_0}{q_0}\alpha,\quad z=\chi,
\end{equation}
where $R_0$ denotes the major radius of the tokamak. 
The equilibrium magnetic field can thus be written as $\bm B = B_0\grad x\times \grad y$ and the Jacobian of the normalized, field aligned, coordinate system writes
\begin{equation}
J_{xyz} = (\grad z \cdot \grad x \times \grad y)^{-1} = B_0/(\grad z \cdot \bm B) = (\hat B b_z)^{-1},
\end{equation}
where $\hat B(z) = B/B_0$ is the normalized magnetic field amplitude and $b_z= \grad z \cdot \bvec$.
In the field aligned coordinate system, the parallel gradient operator writes
\begin{equation}
    \nabla_\parallel= \frac{1}{J_{xyz}\hat B}\ddz.
\end{equation}
The perpendicular nonlinear term in Eq. \eqref{eq:gyboeq} can be expressed as
\begin{equation}
    (\bvec\times \grad f_1)\cdot\grad f_2 = \hat B^{-1}\squareparenthesis{\Gamma_1 \{f_1,f_2\}_{xy} + \Gamma_2 \{f_1,f_2\}_{xz} + \Gamma_3 \{f_1,f_2\}_{yz}},
\end{equation}
where $f_1$ and $f_2$ are two generic spatially dependent fields and $\{f_1,f_2\}_{ij}=\partial_i f_1 \partial_j f_2 - \partial_j f_1 \partial_i f_2$ is the Poisson Bracket operator and we introduce the geometric factors 
\begin{align}
    \Gamma_1&=g^{xx}g^{yy} - g^{xy}g^{xy},\nonumber\\
    \Gamma_2&=g^{xx}g^{yz} - g^{xy}g^{xz},\nonumber\\
    \Gamma_3&=g^{xy}g^{yz} - g^{yy}g^{xz}\nonumber,
\end{align}
being $g^{ij}(z) = \grad i \cdot \grad j$ the metric coefficients for $i,j=x,y,z$.
\par
In this work we consider the $s-\alpha$ magnetic equilibrium model, which assumes circular concentric flux surfaces using a first order approximation with respect to the inverse aspect ratio $\varepsilon_0=r_0/R_0$ \citep{Connor1978ShearModes}.
This includes a constant magnetic shear $\hat s$, such that the safety factor is expressed as $q(x)=q_0(1+x\hat s)$.
The ballooning angle $\chi$ is approximated by the poloidal angle $\theta$, which yields $z=\theta$.
Neglecting the MHD equilibrium pressure effects, the metric coefficients are given by
\begin{equation}
\begin{pmatrix}
g_{xx} & g_{xy} & g_{xz}\\
g_{yx} & g_{yy} & g_{yz}\\
g_{zx} & g_{zy} & g_{zz}
\end{pmatrix}
=
\begin{pmatrix}
1       & \hat sz       & 0\\
\hat sz & 1+(\hat sz)^2     & \varepsilon^{-1}\\
0       & \varepsilon^{-1} & \varepsilon^{-2}
\end{pmatrix},
\end{equation}
where $\varepsilon=r_0(1+x)/R_0$.
The $s-\alpha$ geometry retains finite aspect ratio term in the expression of the magnetic field amplitude $\hat B = 1/(1+\varepsilon \cos z)$, which yields the derivatives $\ddx \ln B = -\cos(z)\hat B/R_0$, $\ddy \ln B = 0$, $\ddz \ln B = \epsilon\sin(z)\hat B$,
and the Jacobian takes the form $J_0=q_0 B_0/B$.
While the $s-\alpha$ model is known to present inconsistencies in the $\varepsilon$ ordering \citep{Lapillonne2009ClarificationsTurbulence}, we consider it here with the purpose of establishing direct comparisons with previous literature results \citep{Lin1999EffectsTransport,Dimits2000ComparisonsSimulations,Frei2023Moment-basedModel}.
\subsection{Scale separation}
We consider a scale separation between perpendicular and parallel fluctuations, ordering them such that $k_{x}\rho_{s} \sim k_{y}\rho_{s} \sim k_z R_0$, where $\rho_{s} = m_i c_{s}/(q B_0)$ is the reference sound Larmor radius and $c_{s0}=\sqrt{T_e/m_i}$ the reference sound speed.
Assuming $\rho_{s0}/R_0 \sim \Delta$, this implies that $k_z/k_{x,y}\sim \Delta$, which is valid for fluctuations in typical conditions of the core and pedestal of JET and ITER \citep{Giroud2015ProgressWall}.
Consequently, we write the magnetic curvature operator as $(\bvec\times\grad \ln B)\cdot \grad = \Gamma_1\hat B^{-1} \mathcal C_{xy}$, with
\begin{equation}
    \mathcal C_{xy} = -\squareparenthesis{\partial_y\ln B + \frac{\Gamma_2}{\Gamma_1}\partial_z\ln B} \frac{\partial}{\partial x} +
    \squareparenthesis{\partial_x\ln B - \frac{\Gamma_3}{\Gamma_1}\partial_z\ln B} \frac{\partial}{\partial y},
    \label{eq:magn_curv_op}
\end{equation}
where we neglect terms related to parallel derivatives.
The perpendicular nonlinear term present in \refeq{eq:gyboeq} can be simplified by using the same ordering, that is
\begin{align}
   (\bvec\times\grad\phibar)\cdot\grad g = \Gamma_1\hat B^{-1}\{\phibar,g\}_{xy}.
\end{align}
Finally, we note that the parallel nonlinear term can be neglected since
\begin{equation}
    (q/m) b_z \ddz \phibar \ddvpar g \sim \phibar g/(B_0R_0\rho_{s0})\ll \Delta.
\end{equation}
Within these assumptions and considering radially dependent density and temperature equilibrium profiles, the GK Boltzmann equation, Eq. \eqref{eq:gyboeq}, writes
\begin{align}
    \ddt g &+ \frac{\hat B}{B} \{\phibar,g\}_{xy}+ \frac{\vtha^2}{2\Omega_a}\roundparenthesis{2s_\parallel^2+w_\perp}\hat B \mathcal C_{xy}h \nonumber\\
    &+ \frac{\hat B}{B} \squareparenthesis{\ddx\ln N + \roundparenthesis{s_\parallel^2+w_\perp-\frac{3}{2}}\ddx \ln T}F_0\ddy\phibar\nonumber\\
    &+ \frac{v_{thi}}{2J_{xyz}\hat B}\squareparenthesis{2s_\parallel \ddz h - w_\perp \ddz\ln B\ddspar h} = C_{ii},
    \label{eq:3d_gyboeq_dim}
\end{align}
where we introduce the dimensionless velocity coordinates $s_\parallel = \vpar/v_{thi}$ and $w_\perp=\mu B/T_i$, and use the relation $\Gamma_1=\hat B^2$.
\par
\begin{table}
    \centering
    \begin{tabular}{l r l| l r l}
    Parallel velocity     & $s_\parallel$ & $= \vpar^{ph}/v_{th i}$  
    &
    Perpendicular velocity     & $w_\perp$ & $= \mu^{ph} B/T_i$ 
    \\
    Perpendicular spatial scales & $k_{x,y}$ & $= k_{x,y}^{ph}\rho_{s}$
    &
    Normalized time & $t$ & $= t^{ph} c_{s}/R$
    \\
    Density gradient & $\kappa_T$ & $= R_0/L_{Ti}$ 
    &
    Temperature gradient & $\kappa_N$ & $= R_0/L_{N}$ 
    \\
    Ion charge number & $q$ & $= q_i^{ph}/e$ 
    &
    Temperature ratio & $\tau$ & $= T_i/T_{e}$
    \\
    Particle mass ratio & $\sigma$ & $= \sqrt{m_i/m_e}$ 
    &
    Distribution function & $f$ & $= f^{ph}/F_0$ 
    \\
    Electrostatic potential & $\phi$ & $= e\phi^{ph}/T_{e} $
    &
    Collision frequency & $\nu$ & $= \nu^{ph}/\nu_{ii}$
    \end{tabular}
    \caption{Dimensionless variables used in the gyromoment model. 
    For a dimensionless variable $A$, its equivalent in physical units is explicitly denoted as $A^{ph}$.}
    \label{tab:dimensionless_units}
\end{table}
In the rest of this work, we express all quantities in dimensionless units according the normalization presented in Table \ref{tab:dimensionless_units}\reviewii{, except for the figure labels where we use physical units}.
In the flux tube limit, we impose constant background gradient profiles, assuming $A(\bm x)\sim A(0)=A_0$ and $|\grad A| \sim |A_0/L_A|$ for a generic background quantity $A$.
Finally, the dimensionless, scale separated, flux tube GK Boltzmann equation writes
\begin{align}
    \ddt g &+ \{\phibar,g\}_{xy}+ \frac{\tau }{q}\roundparenthesis{2 s_\parallel^2+w_\perp } \mathcal C_{xy} h 
    +\squareparenthesis{\kappa_N + \roundparenthesis{s_\parallel^2+w_\perp-\frac{3}{2}} \kappa_T}\ddy \phibar
    \nonumber\\
    &+ \frac{\sqrt{2}}{2}\sqrt{\tau}\frac{1}{J_{xyz}\hat B}\squareparenthesis{2s_\parallel \ddz h - w_\perp\ddz\ln B\ddspar h} 
    = C_{ii},
    \label{eq:3d_gyboeq_nondim}
\end{align}
introducing $\kappa_N={R_0}/{L_N}$ the normalized background density gradient, $\kappa_T={R_0}/{L_T}$ the normalized background temperature gradient, $\tau$ the ion-electron temperature ratio, $q$ the ion charge number and $\sigma$ the electron-ion mass ratio.
\subsection{Nonlinear Hermite-Laguerre-Fourier pseudo spectral formulation}
We project the dimensionless GK perturbed distribution function onto a Hermite and Laguerre polynomial basis \citep{Grant1967Fourier-HermiteLimit,Madsen2013Full-FModel,Manas2017ImpactSimulations,Adkins2018ASpace,Staebler2023AEquations}
\begin{equation}
    g(x,y,z,s_\parallel,w_\perp,t) = \sum_{p=0}^\infty\sum_{j=0}^\infty M^{pj}(x,y,z,t) H_p(s_\parallel)L_j(w_\perp),
\end{equation}
where we introduce $M^{pj}(x,y,z,t)$, the ion GM of order $(p,j)$.
The Hermite polynomial of order $p$ is defined as
\begin{equation}
    H_p(s_\parallel)=\frac{(-1)^p }{\sqrt{2^p p!}} e^{s_\parallel^2} \frac{ d^p}{ ds_\parallel^p} e^{-s_\parallel^2},
    \label{eq:hermite}
\end{equation}
which is a normalized version of the physicist's Hermite polynomials such that $\int_{-\infty}^\infty \mathrm d s_\parallel H_p H_{p'} e^{-s_\parallel ^2}=\sqrt{\pi}\delta_{pp'}$ where $\delta_{pp'}$ denotes the Kronecker delta.
With this definition, useful Hermite product and derivation identities follow, such as $s_\parallel H_p = \sqrt{(p+1)/2}H_{p+1} + \sqrt{p/2}H_{p-1}$ and $\partial_{s\parallel} H_p =\sqrt{2p}H_{p-1}$.
The Laguerre polynomial of order $j$ is expressed as 
\begin{equation}
     L_j(w_\perp)=\frac{e^{w_\perp}}{j!}\frac{ d^j}{ dw_\perp^j}w_\perp^j e^{-w_\perp}
    \label{eq:laguerre}
\end{equation}
and satisfies the orthogonality relation $\int_{0}^\infty \mathrm d w_\perp L_j L_{j'} e^{-w_\perp}=\delta_{jj'}$.
The Laguerre polynomials also satisfy the product identity $w_\perp L_j = (2j+1)L_j-(j+1)L_{j+1}-jL_{j-1}$ \citep{Gradshteyn2014TableProducts}.
\par
By using the orthogonality relations, the GMs are obtained from the distribution function as
\begin{equation}
    M^{pj}(x,y,z,t) = \iint\mathrm dw_\perp \mathrm ds_\parallel g(x,y,z,s_\parallel,w_\perp,t) H_p(s_\parallel) L_j(w_\perp).
\end{equation}
\par
The flux tube model considers the $\grad x$ and $\grad y$ directions as periodic, which is valid as long as the domain size is larger than the \reviewiii{perpendicular} correlation length of the turbulent eddies \citep{Ball2020}. 
It is thus convenient to express our fields in terms of $(k_x,k_y)$ Fourier modes, i.e.
\begin{equation}
    N^{pj}(k_x,k_y,z,t) = \iint\mathrm{d}x \mathrm{d}y M^{pj}(x,y,z,t) e^{-ik_x x-ik_y y }.
\end{equation}
It follows that the Fourier representation of the gyroaveraged electrostatic potential yields
\begin{equation}
    J_0\left(\sqrt{l w_\perp}\right)\phi(k_x,k_y,z,t) = \iint\mathrm{d}x \mathrm{d}y \phibar(\bm x,t) e^{-ik_x x-ik_y y },
\end{equation}
with $J_0$ the Bessel function of the first kind, which can be expressed in terms of Laguerre polynomials as
\begin{equation}
    J_0\left(\sqrt{l w_\perp}\right) = \sum_{n=0}^\infty \kernel_n(l)L_n(w_\perp),
\label{eq:bess_lag}
\end{equation}
where $\kernel_n(l)=l^n e^{-l}/n!$, $l(k_x,k_y,z)=\tau \kperp^2/2$ and the perpendicular wavenumber $\kperp^2 = g^{xx}k_x^2 + 2 g^{xy}k_x k_y + g^{yy} k_y^2$ \citep{Frei2020}.
The projection of \refeq{eq:3d_gyboeq_nondim} onto the Hermite-Laguerre basis yields the gyromoment nonlinear hierarchy in a flux tube configuration,
\begin{equation}
    \ddt N^{pj} + \mathcal S^{pj} + \mathcal M_{\perp}^{pj} + \mathcal M_{\parallel}^{pj} + \mathcal D_{N}^{pj} + \mathcal D_{T}^{pj} = \mathcal C_{ii}^{pj}.
    \label{eq:moment_hierarchy}
\end{equation}
In \refeq{eq:moment_hierarchy}, the perpendicular magnetic term, related to the curvature and gradient drifts, writes
\begin{align}
    \mathcal M_{\perp}^{pj} &= \frac{\tau}{q} \mathcal C_{k_x k_y} \squareparenthesis{\sqrt{(p+1)(p+2)} n^{p+2,j} + (2p+1)n^{pj} + \sqrt{p(p-1)}n^{p-2,j}}
    \nonumber\\&
    + \frac{\tau}{q} \mathcal C_{k_x k_y} \squareparenthesis{(2j+1)n^{pj} - (j+1)n^{p,j+1}-jn^{p,j-1}},
    \label{eq:Mperpapj}
\end{align}
\reviewii{with $C_{k_x k_y}$ the magnetic curvature operator of Eq. \eqref{eq:magn_curv_op} expressed in Fourier space, }while the parallel magnetic term, related to the Landau damping and the mirror force, is expressed as
\begin{align}
    \mathcal M_{\parallel}^{pj} =& 
    \frac{\hat B^{-1}}{J_{xyz}} \sqrt{\tau}\left\{\ddz \aleph^{p\pm1,j}\right.
    - \ddz\ln B \left[(j+1)\aleph^{p\pm1,j}-j\aleph^{p\pm1,j-1}\right]
    \nonumber\\&
    \qquad\left. +\ddz\ln B\sqrt{p}\squareparenthesis{(2j+1)n^{p-1,j} -(j+1)n^{p-1,j+1} - jn^{p-1,j-1}}\right\}
    \label{eq:Mparapj}
\end{align}
with $\aleph^{p\pm1,j}=\sqrt{p+1} n^{p+1,j} + \sqrt{p} n^{p-1,j}$. 
The background gradient drift terms are
\begin{align}
    \mathcal D_{N}^{pj} = ik_y\kappa_N\kernel_j\phi\delta_{p0},
    \label{eq:DNapj}
\end{align}
for the density and
\begin{align}
    \mathcal D_{T}^{pj} = ik_y\phi\kappa_T \left\{\kernel_j\squareparenthesis{\frac{1}{\sqrt{2}}\delta_{p2} -\delta_{p0}}+ \squareparenthesis{(2j+1)\kernel_j-(j+1)\kernel_{j+1}-j\kernel_{j-1}}\delta_{p0}\right\},
    \label{eq:DTapj}
\end{align}
for the temperature.
In Eqs. \eqref{eq:Mperpapj}, \eqref{eq:Mparapj} and \eqref{eq:DTapj}, we also introduce the non-adiabatic ion GM, $n^{pj}(\bm k,t)=N^{pj}+q \phi \kernel_j\delta_{p0}/\tau$.\\
The nonlinear term related to the $\bm E\times \bm B$ drift is expressed as
\begin{equation}
    \mathcal{S}^{pj} =  \sum_{n=0}^{\infty}\curlyparenthesis{\sum_{s=0}^{n+j}d_{njs} N^{ps},\kernel_n\phi}_{xy},
    \label{eq:sapj}
\end{equation}
where we use the Bessel-Laguerre decomposition, \refeq{eq:bess_lag}, and we express the product of two Laguerre polynomials as a sum of single polynomials using the identity
\begin{equation}
L_jL_n=\sum_{s=0}^{n+j}d_{njs}L_s
\label{eq:lagprod}
\end{equation}
with
\begin{equation}
    d_{njs} = \sum_{n_1=0}^n\sum_{j_1=0}^j\sum_{s_1=0}^s \frac{(-1)^{n_1+j_1+s_1}}{n_1!j_1!s_1!}\binom{n}{n_1}\binom{j}{j_1}\binom{s}{s_1}.
    \label{eq:dnjs}
\end{equation}
Finally, using an adiabatic electron response, we close our system with the dimensionless GK Poisson equation in Fourier space, i.e.
\begin{equation}
     \left[ 1 + \frac{q^2}{\tau}\roundparenthesis{1-\sum_{n=0}^{\infty}\kernel^2_n}\right]\phi - \langle \phi \rangle_{yz} = q\sum_{n=0}^{\infty}\kernel_n N^{0n},
    \label{eq:poisson_moments_adiabe}
\end{equation}
where $\langle \phi \rangle_{yz}$ is the flux surface average of $\phi$, namely
\begin{equation}
    \langle \phi \rangle_{yz} = \frac{1}{\int\dz J_{xyz}}\int\dz J_{xyz}\phi(k_x,k_y,z,t)\delta_{k_y0}.
\end{equation}
\reviewi{While the adiabatic electron model considered here serves as a valuable tool for comparison with previous work, it is important to note its inherent limitations. 
In fact, evolving the electrons GK equation for the CBC leads to complex small scales phenomena resulting generally, in an increase of linear growth rate and saturated heat flux level due to the electron temperature gradient (ETG) instability \citep{Neiser2019GyrokineticPlasmas,Hardman2022ExtendedElectrons}.}
%
\par
\modifiv{To model collisions, we consider the GK Landau operator, which considers the linear limit of exact GK Coulomb collisions, along with the GK Sugama collision model \citep{Sugama2009LinearizedEquations} and the GK Dougherty model \citep{Dougherty1964}.
The details of these operators and their projection onto the Hermite-Laguerre basis can be found in \cite{Frei2022LocalMode}. 
For an overview of the qualitative differences between the aforementioned collision operators we also refer to \cite{Hoffmann2023GyrokineticOperators}.
We set the intensity of the collisions through the nondimensional parameter $\nu$, normalized by the ion-ion collision frequency
\begin{equation}
    \nu_{ii} = \frac{4\sqrt{\pi}}{3}\frac{R_0 N e^4 \ln \Lambda}{c_s m_i^{1/2}T_i^{3/2}},
\end{equation}
where $\ln \Lambda$ is the Coulomb Logarithm.
}
\subsection{Numerical approach}
To solve numerically the GM hierarchy, \refeq{eq:moment_hierarchy}, we evolve a finite set of ion Hermite-Laguerre-Fourier modes, $N^{pj}(k_x,k_y,z,t)$, with $0\leq p \leq P$ and $0\leq j \leq J$. 
We label a finite set of GMs by the pair $(P,J)$  where $P$ and $J$ represent the maximal polynomial degree of the considered Hermite and Laguerre basis, respectively.
For the time integration, we use a standard explicit fourth-order Runge-Kutta time-stepping scheme.
\subsubsection{Perpendicular spatial discretization}
The Fourier modes are defined for the wavenumbers $k_x = 2\pi m/L_x$, with $-N_{x}/2 +1\leq m \leq N_{x}/2$, and $k_y=2\pi n/L_y$, with $0\leq n \leq N_{y}/2$ (we exploit the reality condition) being $L_x$ and $L_y$ the perpendicular dimensions of the flux tube.
The nonlinear term, Eq. \eqref{eq:sapj}, is evaluated using a pseudo-spectral method on each perpendicular plane.
More precisely, the spatial derivatives, contained in the Poisson bracket, are obtained in the real space using a backward fast Fourier transform \citep{FFTW05}.
The fields are then multiplied in real space and the result transformed back to Fourier space using a forward fast Fourier transform, including a 2/3 anti-aliasing filter \citep{Orszag1971}.
To prevent energy pile-up due to the turbulent cascade, we include numerical diffusion of the form $\mu_d(ik/k_{max})^4$, where $k_{max}$ is the largest non-aliased allowed wavelength in the simulation and $\mu_d$ a tunable parameter, set usually to $\mu_d =0.5$.
\subsubsection{Parallel spatial discretization}
The parallel direction is discretized using a regular grid of $N_z$ points with spacing $\Delta z=2\pi/(N_z-1)$, such that the grid points are $z_l = l \Delta z - \pi$  for $0\leq l \leq N_z-1$.
This constraints our spatial domain to a single poloidal turn around the flux surface (the effect of using a larger number of poloidal turns is described by \cite{Ball2020} and \cite{Volcokas2023UltraTokamaks}).
For evaluating the parallel derivative coming from the Landau damping term in Eq. \eqref{eq:Mparapj}, we use a four-point, fourth-order, centered finite differences scheme.
\par
To account for the effect of magnetic shear, we adopt a "twist-and-shift" condition for the parallel boundary conditions, which imposes the coupling between $k_x$ and $k_y$ modes, i.e.
\begin{equation}
    A(k_x,k_y,z) = A(k_x+2\pi \hat s k_y,k_y,z+2\pi),
\end{equation}
with $A$ a generic spatially dependent field.
For more details on the twist and shift boundary conditions, we refer to the works of \cite{Beer1995} and \cite{Ball2020}.
We also add a fourth-order parallel numerical diffusion term of the form $\mu_z/\Delta z^4 \partial_z^4$, with $\mu_z$ a tunable parameter, which prevents \reviewi{the decoupling between odd and even points} 
in the $z$ direction \citep{Paruta2018} and \modifiii{helps to smooth out the oscillations due to the Dirichlet boundary condition applied at the end of the ballooning angle domain}. 
We note that a staggered grid representation can also be implemented to avoid the checkerboard pattern, as it is done for Braginskii's equations in \cite{Paruta2018}.

%
\subsubsection{Velocity space discretization and closure}
When a finite set of $(P,J)$ GMs is considered, $(P+1)\times(J+1)$ coupled equations are solved.
In order to close the system, the GMs with degree higher than $P$ or $J$ appearing in Eqs. \eqref{eq:Mperpapj}, \eqref{eq:Mparapj}, and \eqref{eq:sapj}, are assumed to vanish using a truncation closure, i.e. $N^{pj}=0$ for $p > P$ or $j > J$. 
The truncation closure is the most straightforward to apply and shows good results in \cite{Frei2022LocalMode}, \cite{Frei2023Moment-basedModel} and \cite{Hoffmann2023GyrokineticOperators}.
We must note that, in collisionless simulations, the Hermite parallel coupling in Eq. \eqref{eq:Mparapj}, combined with the closure by truncation, may lead to recurrence effects, which yields a nonphysical energy transfer from the highest to the lowest Hermite modes.
When considering the collisionless limit, we avoid recurrence effects by adding a numerical diffusive term in the velocity space,
\begin{align}
D_v^{pj} =& -\eta_{v} (p + 2j) N^{pj} \nonumber\\
&+\eta_{v} \squareparenthesis{ N^{10}\delta_{p1}\delta_{j0} - \frac{2}{3}\roundparenthesis{\sqrt{2}N^{01}-N^{20}}\delta_{p2}\delta_{j0} - \frac{2}{3}\roundparenthesis{\sqrt{2}N^{20}-N^{01}}\delta_{p0}\delta_{j1}},
\end{align}
with $\eta_v$ a tunable parameter.
We note that this term, similar to a Dougherty collision operator, has the advantage of conserving mass, momentum, and energy.
In this work, we ensure that this artificial term does not impact our results by comparing them at different values of $\eta_{v}$ (see Sec. \ref{dimits_convergence_study}).
While a generalization of the Braginskii's equation closure to an arbitrary number of moments is still an open issue, more sophisticated closure models such as the semi-collisional closure proposed by \cite{Zocco2011}, \cite{Loureiro2016Viriato:Dynamics} exist and might reduce the appearance of recurrence effects without introducing an artificial dissipation term.
One can also mention the work of \cite{Shukla2022AGeometry} which presents a method for formulating closures that learn from kinetic simulation data.

\section{Cyclone base case convergence study}
\label{sec:CBC_benchmark}
We present a benchmark and a convergence analysis of the GM approach in the collisionless CBC \citep{Lin1999EffectsTransport,Dimits2000ComparisonsSimulations}. 
We focus on the ITG linear growth rate and the nonlinear saturated heat flux. 
We compare our findings with the results obtained from previous work \citep{Dimits2000ComparisonsSimulations}, as well as simulations carried out by using the continuum code GENE \citep{Jenko2000}\reviewi{, also considering an adiabatic electron response}.
We also investigate the convergence of the GM approach by varying the number of evolved moments.
\par
Unless stated otherwise, we employ the conventional CBC parameters: $\kappa_T=6.96$, $\kappa_N=2.22$, $q_0=1.4$, $\epsilon=0.18$ and $\hat s = 0.8$.
Regarding the resolution, our linear scan encompasses $N_x=8$ coupled $k_x$ modes and $N_z=24$ points in the parallel direction.
For the nonlinear simulations, we set $L_x/\rho_s=120$, $L_y/\rho_s=120$, and $L_z/R_0=2\pi$, with corresponding grid dimensions $N_x=128$, $N_y=64$, and $N_z=24$ in both the GENE and GM codes.
Similarly, a fourth-order parallel dissipation term with $\mu_z=0.2$ is used in both codes \citep{Pueschel2010OnMicroturbulence}.
\modifii{For GENE simulations, we use the standard velocity domain, $L_{\vpar}\times L_\mu =  9\times 3$, where $L_{\vpar}$ and $L_{\mu}$ are the dimensions of the velocity domain along the parallel velocity and magnetic moment directions, respectively.}
Finally, the velocity fourth-order dissipation parameter of GENE is set to $\nu_v=0.2$ \citep{Pueschel2010OnMicroturbulence}, while the GM simulations use the numerical velocity dissipation frequency $\eta_{v}=0.001$.

\subsection{Linear convergence and ion temperature gradient stability threshold}
To evaluate the ITG growth rate, we numerically solve the moment hierarchy equation, Eq. \eqref{eq:moment_hierarchy}, neglecting the nonlinear term in Eq. \eqref{eq:sapj} and we analyze the time evolution of $\phi(t)$ at $k_x=0$ and $z=0$ for various $k_y$ values.
To capture the parallel dynamics and the coupling due to the magnetic geometry, we evolve 8 $k_x$ modes with $\Delta k_x = 2\pi \hat s k_y$.
The growth rate $\gamma$ is determined by fitting the slope of the time evolution of the logarithm of the electrostatic potential amplitude for a given $k_y$ mode.

\begin{figure}
    \centering
    \includegraphics[width=0.49\linewidth]{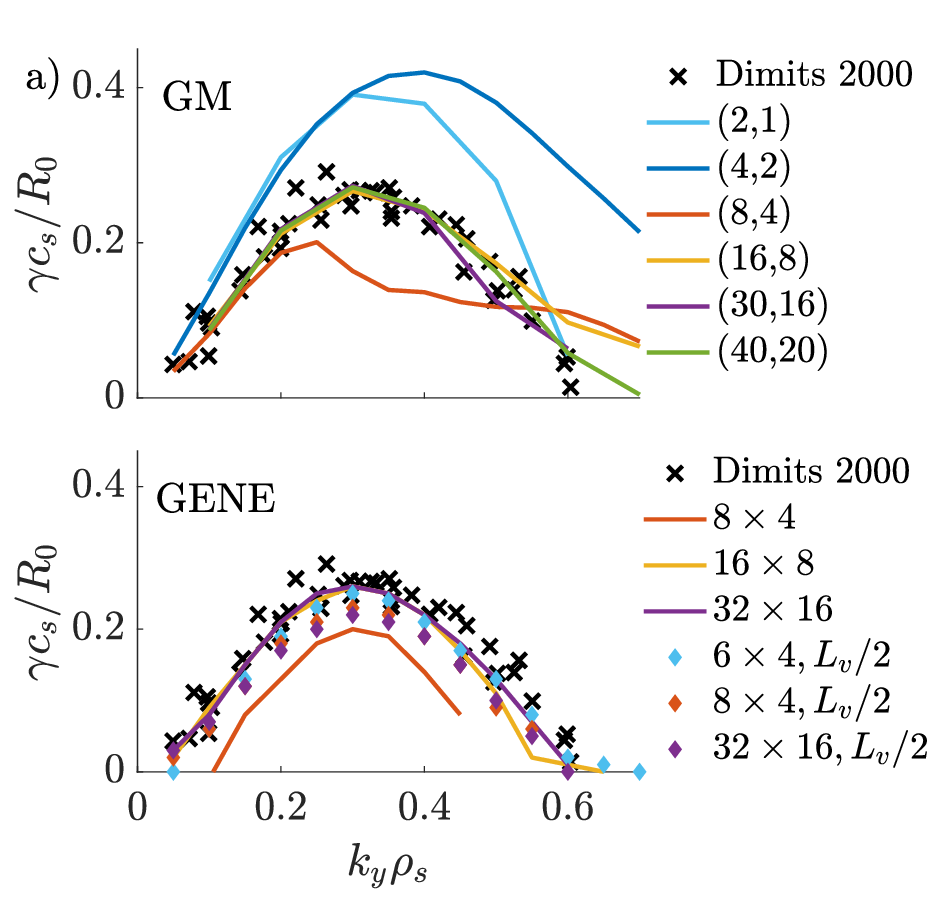}
    \includegraphics[width=0.49\linewidth]{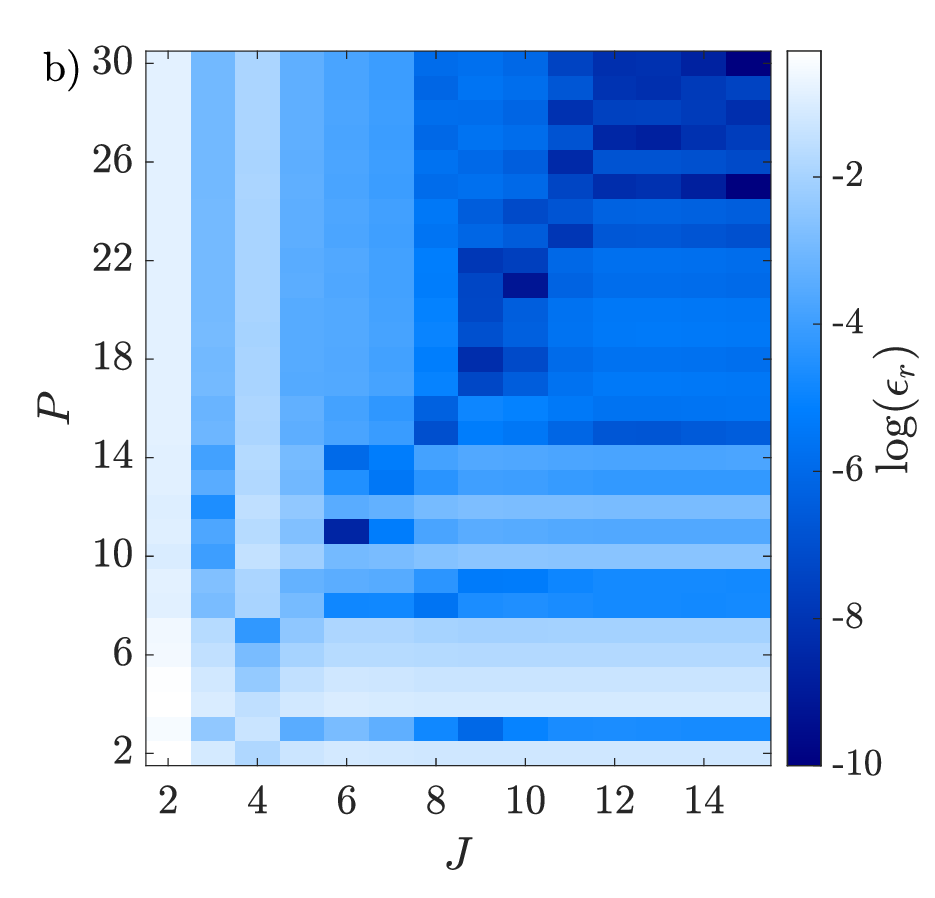}
    \caption{(a) \modifii{CBC linear growth rates, $\gamma$ obtained by using the GM approach (top) and the GENE code (bottom), compared with the results reported in \cite{Dimits2000ComparisonsSimulations} (black crosses). 
    The velocity resolution is scanned by varying the size of the polynomial sets $(P,J)$ for the GM approach and the number of velocity grid points $N_{\vpar}\times N_\mu$ for the GENE code.
    GENE Results with halved velocity domain, $L_{v\parallel}=4.5$ and $L_\mu=1.5$, are denoted as $L_v/2$.}
    (b) Convergence of the relative error $\epsilon_r$ of the CBC linear growth rate obtained with the GM approach at $k_y\rho_s=0.3$. 
    The error is evaluated with respect to the growth rate evaluated with $(P,J)=(60,30)$ GM, $\gamma_{(60,30)}$, namely $\epsilon_r=|\gamma-\gamma_{(60,30)}|/|\gamma_{(60,30)}|$ .}
    \label{fig:31_CBC_linear_convergence}
\end{figure}
Similarly to our previous work on the entropy mode \citep{Hoffmann2023GyrokineticOperators}, we observe a non-monotonic convergence of the GM growth rates with $P$ and $J$.
The GM approach converges to the correct growth rate with a smaller number of polynomials in the long wavelength limit, $k_y\rho_s\ll 1$, than at short wavelengths.
We find that the linear growth rates obtained with the GM approach show good agreement with GENE and \cite{Dimits2000ComparisonsSimulations} results when the polynomial basis is sufficiently large, $(P,J)\gtrsim(16,8)$ (see Fig. \ref{fig:31_CBC_linear_convergence}a).
\\
To analyze the convergence properties of the GM approach in more detail, we show the relative error of the peak growth rate ($k_y\rho_s\simeq 0.3$), $\epsilon_r=|\gamma-\gamma_{(60,30)}|/|\gamma_{(60,30)}|$, in Fig. \ref{fig:31_CBC_linear_convergence}b.
The convergence behavior exhibits typical characteristics of spectral methods, that is an exponential and non-monotonic trend with $P$ and $J$.
It is worth noting that Fig. \ref{fig:31_CBC_linear_convergence}b reveals optimal convergence properties when $P\sim 2J$, which justifies this choice also in previous works \citep{Frei2023Moment-basedModel, Hoffmann2023GyrokineticOperators, Mandell2022GX:Design}.
This is possibly related to the fact that $H_n$ is a Hermite polynomial of degree $n$ in the parallel velocity coordinate, whereas $L_n$ is a Laguerre polynomial of degree $2n$ in the perpendicular velocity coordinate.
We note that, while high $k_y$ modes converge at a slower rate, they have a reduced impact on the saturated transport level.
\reviewi{In fact, according to the mixing length and critical balance estimates (see e.g. \cite{kotschenreuther1995}, \cite{Ricci2006}, \cite{Schekochihin2008GyrokineticSpace}, \cite{Barnes2011CriticallyPlasmas} and \cite{Adkins2023ScaleTurbulence}), the transport of the unstable modes scales as $\gamma/k_y^2$.}

\modifii{
To compare the linear convergence properties between the GM approach and the GENE code, we present a convergence study of the linear growth rate carried out with the GENE code in Fig. \ref{fig:31_CBC_linear_convergence}a.
We observe that the GM approach converges faster than the GENE code for small wavenumbers, which can be attributed to the fluid nature of these modes \reviewi{and to a faster convergence of the Bessel-Laguerre decomposition (see Eq. \eqref{eq:bess_lag}}).
At low velocity resolution, our results show that the GENE code tends to reduce the linear growth rate.
This is in contrast to the GM approach, which overestimates the linear growth rates for both the $(2,1)$ and $(4,2)$ basis.
In addition, Figure \ref{fig:31_CBC_linear_convergence} presents GENE simulations with a reduced velocity domain, $L_{\vpar}\times L_\mu = 4.5\times 1.5$ (denoted $L_v/2$).
Due to the associated decrease of the velocity grid spacing, we observe an improvement of the results.
However, we note that reducing further the velocity domain stabilizes the ITG mode.
Similarly, using a $N_{\vpar} \times N_\mu = 6 \times 4$ grid does not yield unstable growth rates when the standard size of the velocity domain is considered.
}
\subsection{Nonlinear cyclone base case}
We now turn our attention to the nonlinear regime and assess the transport level by examining the normalized ion heat flux $Q_x$. 
\begin{figure}
    \centering
    \includegraphics[width=0.49\linewidth]{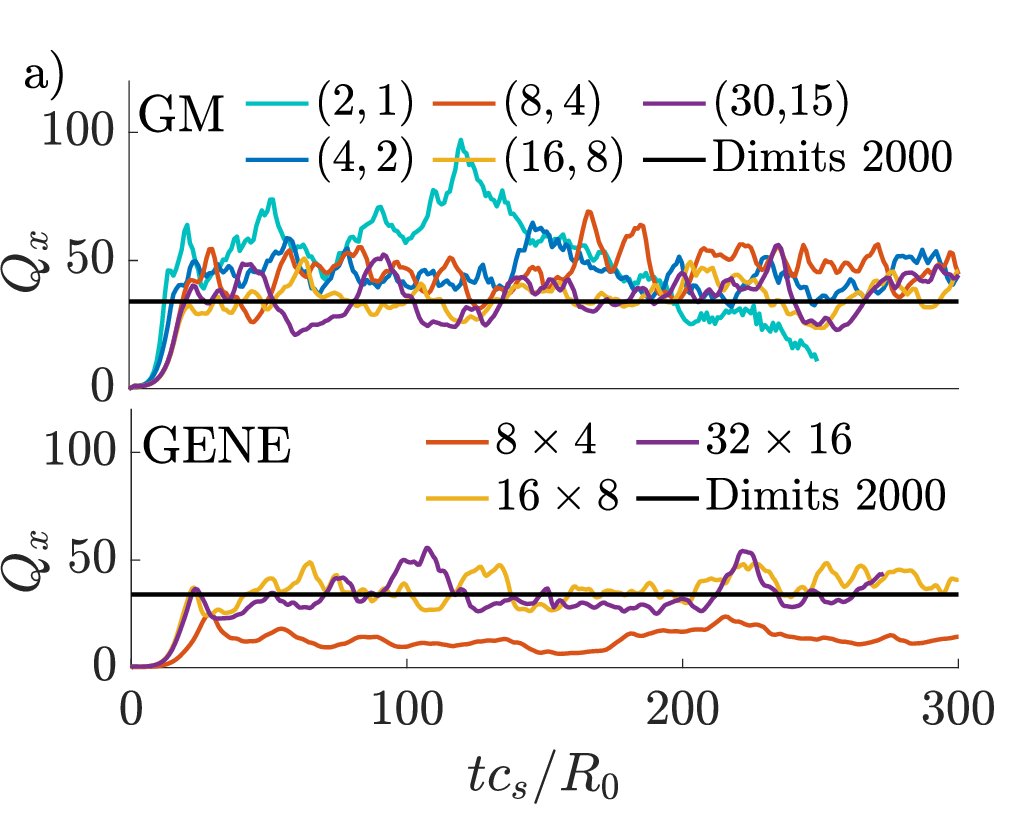}
    \includegraphics[width=0.49\linewidth]{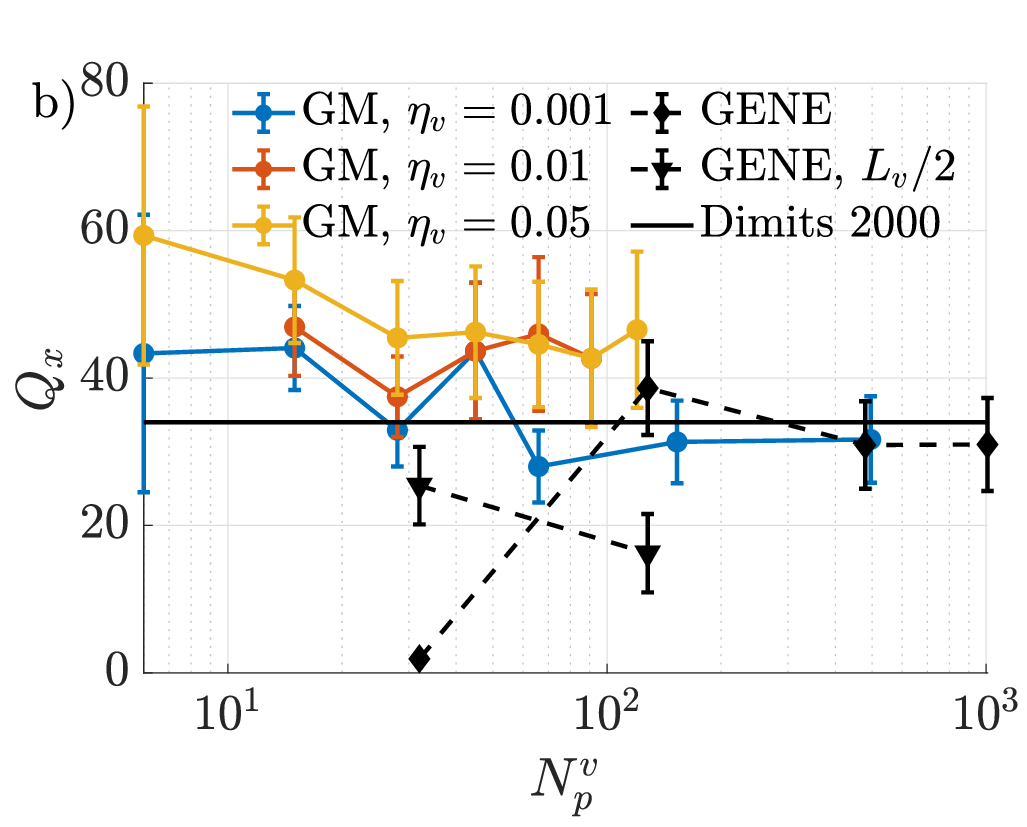}
    \caption{(a) Time traces of the heat flux for the CBC with comparison between the gyromoment approach, GENE and the result of \cite{Dimits2000ComparisonsSimulations}\reviewiii{, $Q_x\approx 35$}.
    Various velocity resolutions are used, in particular different GM sets $(P,J)$ and numbers of points in the GENE velocity grid $N_{\vpar}\times N_\mu$.
    (b) Convergence in the saturated heat flux value with respect to the number of points representing the velocity space $N_p^v=(P+1)\times(J+1)$ for the GM approach and $N_p^v = N_{\vpar}\times N_\mu$ for GENE. 
    The configuration space resolution is $N_x=128$, $N_y=64$ and $N_z=24$ for both the GENE code and the GM approach. The error bars reflect the average fluctuation amplitude around the time-averaged transport value.
    The GENE simulations denoted by $L_v/2$ are obtained with a halved velocity domain size.}
    \label{fig:33_CBC_nonlin_heat_flux}
\end{figure}
Figure \ref{fig:33_CBC_nonlin_heat_flux}a presents the time traces of the heat flux obtained using different GM sets and varying velocity grid resolutions with the GENE code. 
Both codes converge towards the heat flux value obtained by \cite{Dimits2000ComparisonsSimulations}.
This is also shown in Fig. \ref{fig:33_CBC_nonlin_heat_flux}b, where the time-averaged value of the heat flux is shown.
We observe that both GENE and the GM codes yield consistent results, exhibiting similar average values and fluctuation amplitudes.
With the velocity dissipation frequency $\eta_{v}=0.001$, the GM approach demonstrates faster convergence than the GENE code, providing a good estimate of the heat flux with, approximately, 16 Hermite-Laguerre modes.
\modifii{On the other hand, in agreement with the linear results, we observe that the $(4,2)$ GM set overestimates the saturated heat flux level.
We note that the $(2,1)$ GM set fails to saturate, leading to a numerical crash.}
Figure \ref{fig:33_CBC_nonlin_heat_flux}b also presents the convergence behavior for two cases with larger numerical velocity dissipation, namely $\eta_{v}=0.01$ and $\eta_{v}=0.05$. 
In these cases, the moment approach converges more rapidly than with $\eta_{v}=0.001$, but a \reviewii{$30\%$} discrepancy in the converged heat flux value is observed.
\modifii{Since the numerical velocity dissipation term in the GENE code is normalized by the velocity grid size \citep{Pueschel2010OnMicroturbulence}, the incorrect results obtained with the $\Nvpar\times N_\mu=8\times4$ and $16\times8$ resolutions is expected to result from the increased strength of the GENE velocity numerical dissipation.
\modifi{However, the simulations where we reduce by a factor two the size of the velocity domain, yield significant changes in the heat flux values (see Fig. \ref{fig:33_CBC_nonlin_heat_flux}b) and no clear convergence towards the results obtained with the highest resolutions.}}
Thus, we conclude that the GENE code requires, at least, one hundred grid points in the velocity space to provide a good estimate of the transport level.
It is worth noting that, due to recurrence effects, the $(P,J)=(2,1)$ GM basis presents a pile-up of energy at the low-order Hermite-Laguerre modes, yielding an inaccurate transport level.
Alternative closures can potentially prevent this pile up and their use is left for future work.
\par
\begin{figure}
    \centering
    \includegraphics[width=1.0\linewidth]{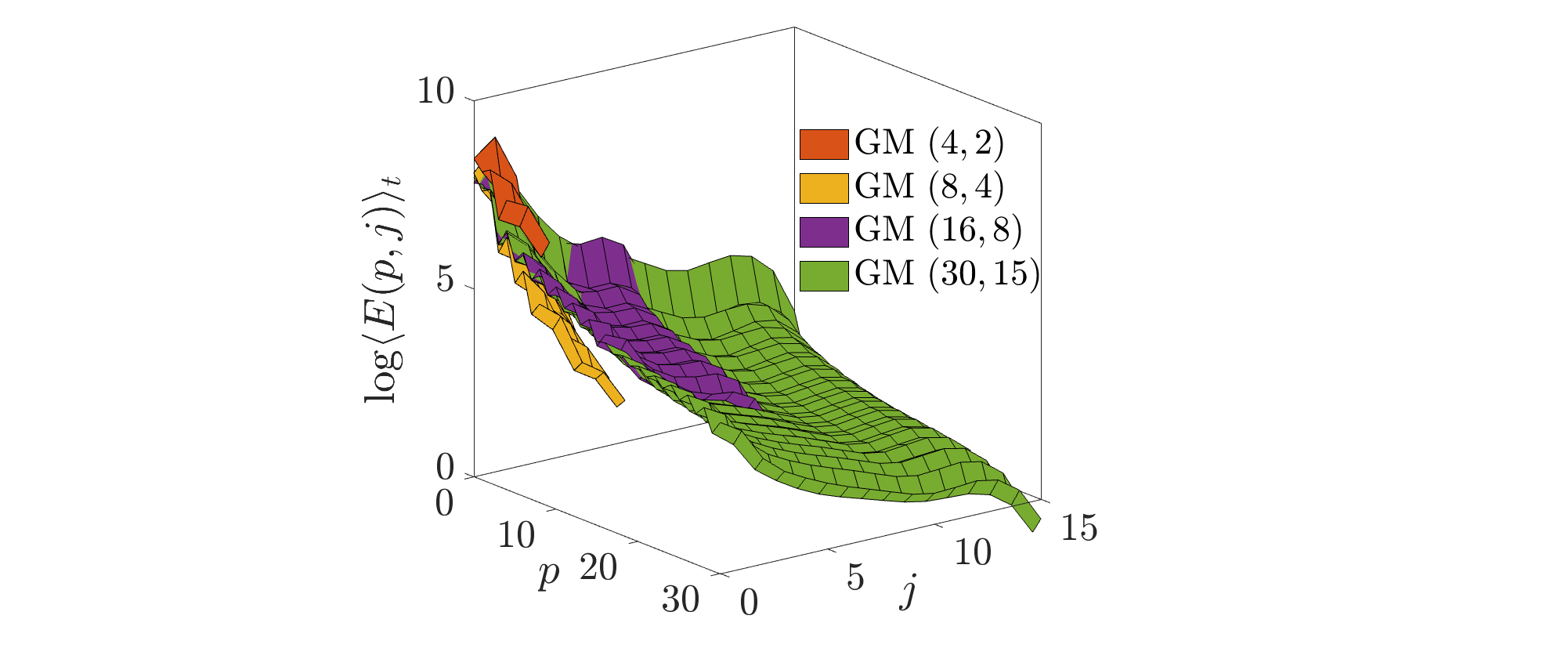}
    \caption{Amplitude of the GM, $E_{pj}=\sum_{k_x,k_y,z}|N^{pj}(k_x,k_y,z)|$, in the CBC, time-averaged during the quasi-steady state.}
    \label{fig:34_CBC_mom_spectrum}
\end{figure}
To further analyze the Hermite-Laguerre representation, we examine the amplitude of the individual GM, defined as $E_{pj}=\sum_{k_x,k_y,z}|N^{pj}(k_x,k_y,z)|$, time-averaged during the quasi-steady saturated state shown in Fig. \ref{fig:34_CBC_mom_spectrum}.
We observe that the truncation closure used in our simulations affects the Hermite and Laguerre modes differently. 
\modifiii{In fact, a plateau followed by a sharp decrease can be observed at higher Laguerre degrees $j$ in both the $(16,8)$ and $(30,15)$ spectra.}
\modifii{On the other hand, a decrease with an approximately constant slope is observed as a function of $p$.}
In more details, the $(4,2)$ GM basis is characterized by a spectrum where the amplitude of the evolved moments is overestimated, although the transport is in good agreement with the converged case. 
The $(8,4)$ basis shows a good agreement in the amplitude of the low-degree GMs whereas the $(16,8)$ simulation shows good agreement with the largest resolution case at all $p$ and $j$.
\reviewi{The reason behing the differences observed between the GM energy spectra in the Hermite and Laguerre directions is twofold. 
First, the GM hierarchy \eqref{eq:moment_hierarchy} couples the Hermite modes mostly through the magnetic curvature and gradient drifts, connecting the $(p,j)$ GM with the $(p\pm2,j)$ GMs, whereas the Laguerre coupling occurs mostly through neighbouring Laguerre modes, i.e. the $(p,j)$ GM is coupled with the $(p,j\pm1)$ GMs. 
This explains the oscillations observed in the Hermite modes ($p$ direction in Fig. \ref{fig:34_CBC_mom_spectrum}).
Second, we recall that the Laguerre polynomials are involved in the Bessel-Laguerre decomposition of Eq. \eqref{eq:bess_lag}, which converges non monotonically with respect to the maximal Laguerre degree, explaining the non monotonic convergence at the highest Laguerre modes in Fig. \ref{fig:34_CBC_mom_spectrum}.}
\par
\reviewiii{We note that the convergence of the nonlinear GM simulations is set by the convergence of the peak linear growth rate of the ITG mode,} which is in agreement with the findings of our previous work \citep{Hoffmann2023GyrokineticOperators}.
In fact, the overestimate of the linear growth rates observed for the $(4,2)$ basis (see Fig. \ref{fig:31_CBC_linear_convergence}a) corresponds to an increased saturated transport value (see Fig. \ref{fig:33_CBC_nonlin_heat_flux}b) and a high amplitude of the GM spectrum on Fig. \ref{fig:34_CBC_mom_spectrum}.
On the other hand, the $(8,4)$ basis presents a reduced amplitude for the high degree GMs.

\section{Collisionless Dimits shift}
\label{sec:dimits_shift}
Building upon the study presented in Section \ref{sec:CBC_benchmark}, we now focus on evaluating the capability of the GM approach to accurately capture the Dimits shift, i.e. the difference between the threshold value of the linear instability and the gradient where a substantial increase of the nonlinear saturated transport level occurs \citep{Dimits2000ComparisonsSimulations}. 
Throughout this section, unless explicitly mentioned, we retain the spatial resolution and numerical dissipation parameters used in Sec. \ref{sec:CBC_benchmark}.

\subsection{Collisionless ITG linear threshold}
In order to investigate the convergence of the Dimits shift, \reviewi{we define the linear instability threshold as the maximum value of the background temperature gradient for which no growth rates exceeding 0.001 can be observed for $0.05\leq k_y \leq 1.0$. 
This criterion is chosen both due to the inherent difficulty in accurately measuring marginal stability and the acknowledgment that turbulence emerging from such low growth rates, requiring thousands of time units to develop, is expected to have a negligible impact on the plasma dynamics}.
\\
\begin{figure}
    \centering
    \includegraphics[width=0.49\linewidth]{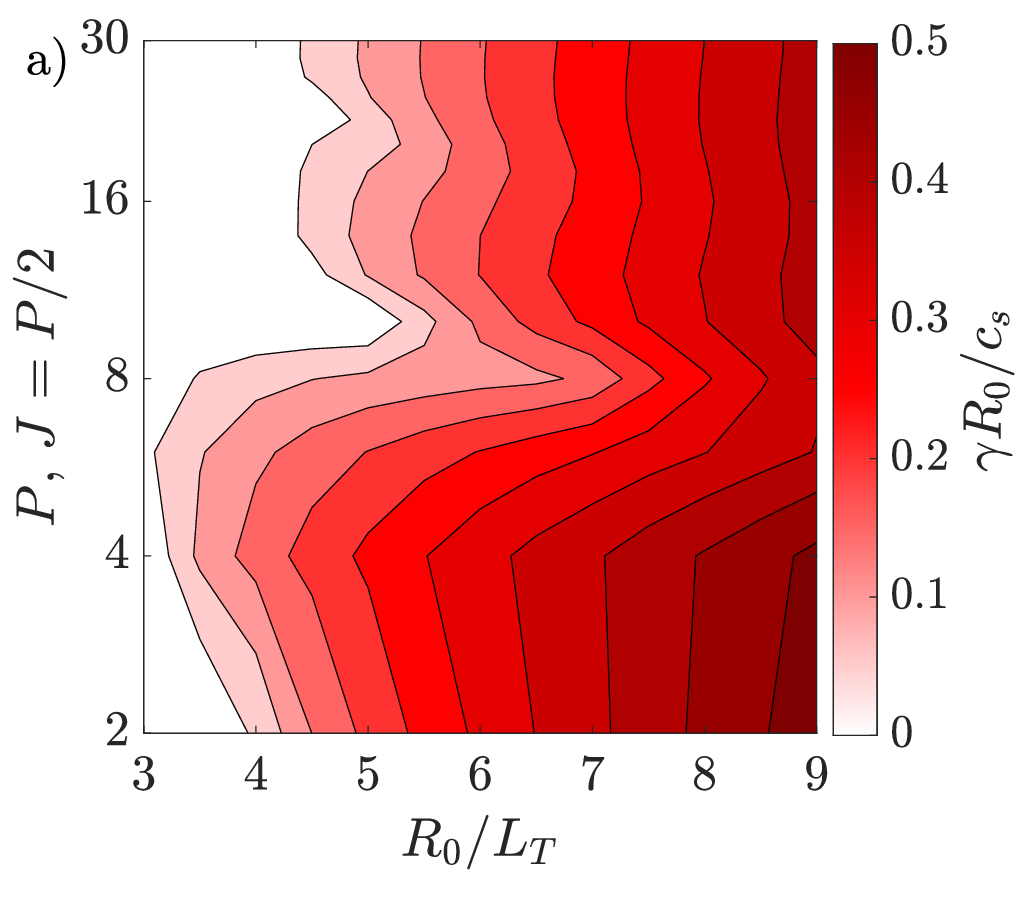}
    \includegraphics[width=0.49\linewidth]{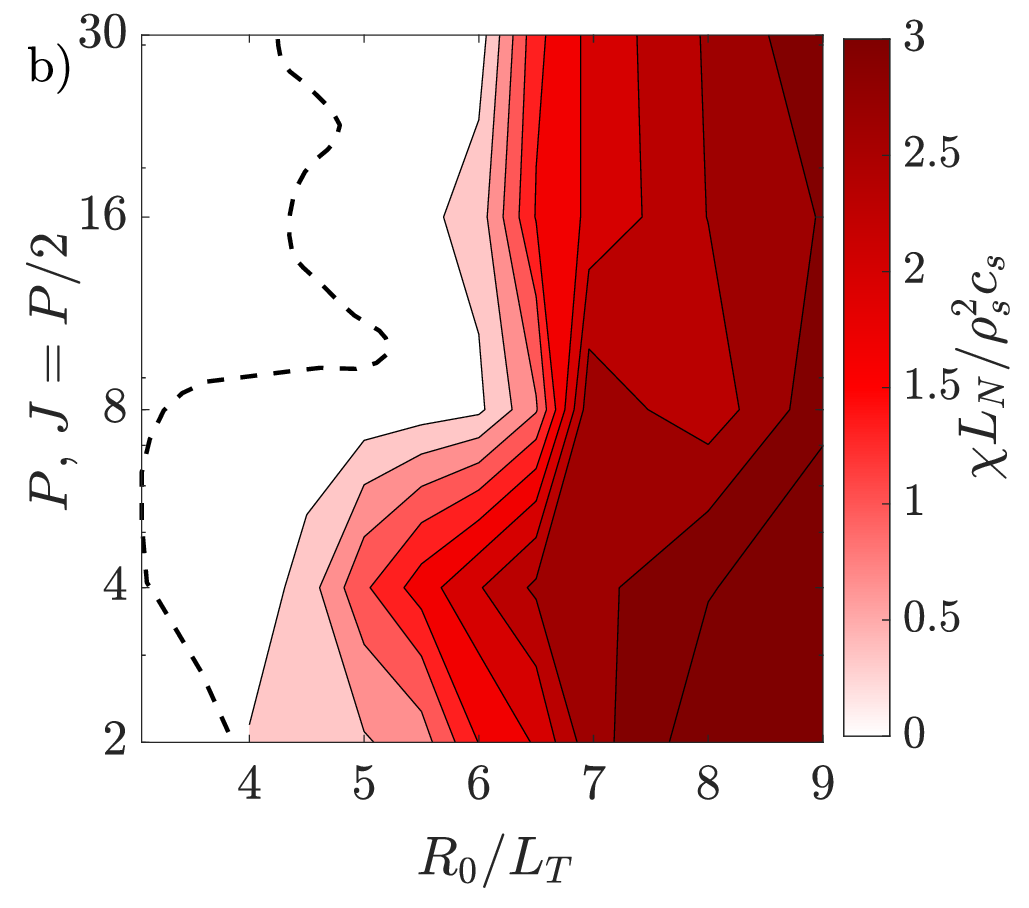}
    \caption{(a) ITG growth rate at $k_y\rho_s=0.3$ for different background temperature value $R_0/L_T$ and $(P,J)$, with $J=P/2$, $R_0/L_N=2.22$ and $\eta_{v}=0.001$.
    (b) Heat diffusivity obtained with the GM approach and the threshold for the ITG linear growth rate (black dashed line). 
    The gap in the $R_0/L_T$ values between the ITG linear threshold and the non-zero $\chi$ values represents the Dimits shift.}
    \label{fig:41_ITG_threshold_and_shift_convergence}
\end{figure}

Figure \ref{fig:41_ITG_threshold_and_shift_convergence}a presents the results of a parameter scan involving the temperature gradient strength $\kT$ and the number of GMs, having fixed $P=2J$, $\eta_{v}=0.001$ and all other parameters as in Sec. \ref{sec:CBC_benchmark}.
Our findings indicate that the GM approach tends to overestimate the ITG threshold when a reduced velocity basis is used, which is in agreement with the observations of \cite{Dimits2000ComparisonsSimulations} for GF codes.
However, for $(P,J)\gtrsim(12,6)$, the GM approach yields the threshold value from GK codes reported by \cite{Dimits2000ComparisonsSimulations}, i.e. $\kappa_T\sim 4$.
Furthermore, convergence is faster when the gradient level is larger, a trend consistent with \cite{Hoffmann2023GyrokineticOperators}.
We recall here that this behavior is a consequence of the relative increase of the importance of gradient drift terms (see Eqs. \eqref{eq:DNapj} and \eqref{eq:DTapj}), compared to the parallel and perpendicular linear terms (ee Eqs. \eqref{eq:Mperpapj} and \eqref{eq:Mparapj}), resulting in reduced coupling of high-degree moments.

\subsection{Dimits shift and nonlinear convergence study}
\label{dimits_convergence_study}
Turning now to the nonlinear dynamics, we perform a set of simulations to investigate the dependence of the heat transport on the temperature gradient values.
We consider the following five sets of GMs: $(P,J)=(2,1)$, $(4,2)$, $(8,4)$, $(16,8)$, and $(30,15)$. 
In addition, we examine the convergence properties of the continuum code GENE by varying the velocity grid resolution, specifically $(N_{v_\parallel},N_\mu)=(8,4)$ and $(16,8)$ with parallel velocity and magnetic moment domain of size $L_{\vpar}\times L_\mu=4.5\times 1.5$. 
We also consider the $(N_{v_\parallel},N_\mu)=(32,16)$ and $(N_{v_\parallel},N_\mu)=(42,24)$ grids in the velocity domain for $L_{\vpar}\times L_\mu=9\times 3$.
We measure the radial heat transport by evaluating the ion heat diffusivity, $\chi=\langle Q_x \rangle_t/(\kappa_T \kappa_N)$, with the heat flux averaged over time during the saturated phase of the nonlinear simulations.

Figure \ref{fig:41_ITG_threshold_and_shift_convergence}b displays the heat diffusivity obtained using the GM approach as a function of the background temperature gradient and the number of evolved Hermite-Laguerre moments. 
By comparing the heat diffusivity with the linear instability threshold determined in the linear case, we also quantify the Dimits shift.
In addition, while the critical gradient is influenced by the number of Hermite-Laguerre moments, the width of the shift is not. 
\reviewi{This provides further support to the observation made in the study of the entropy mode \citep{Hoffmann2023GyrokineticOperators}, which suggests that the constraints on the GM approach convergence are related to the resolution of the primary instability (here the ITG) and not the secondary (the Kelvin-Helmholtz instability) or a possible tertiary one (zonal flows destabilization).}

\begin{figure}
    \centering
    \includegraphics[width=1.0\linewidth]{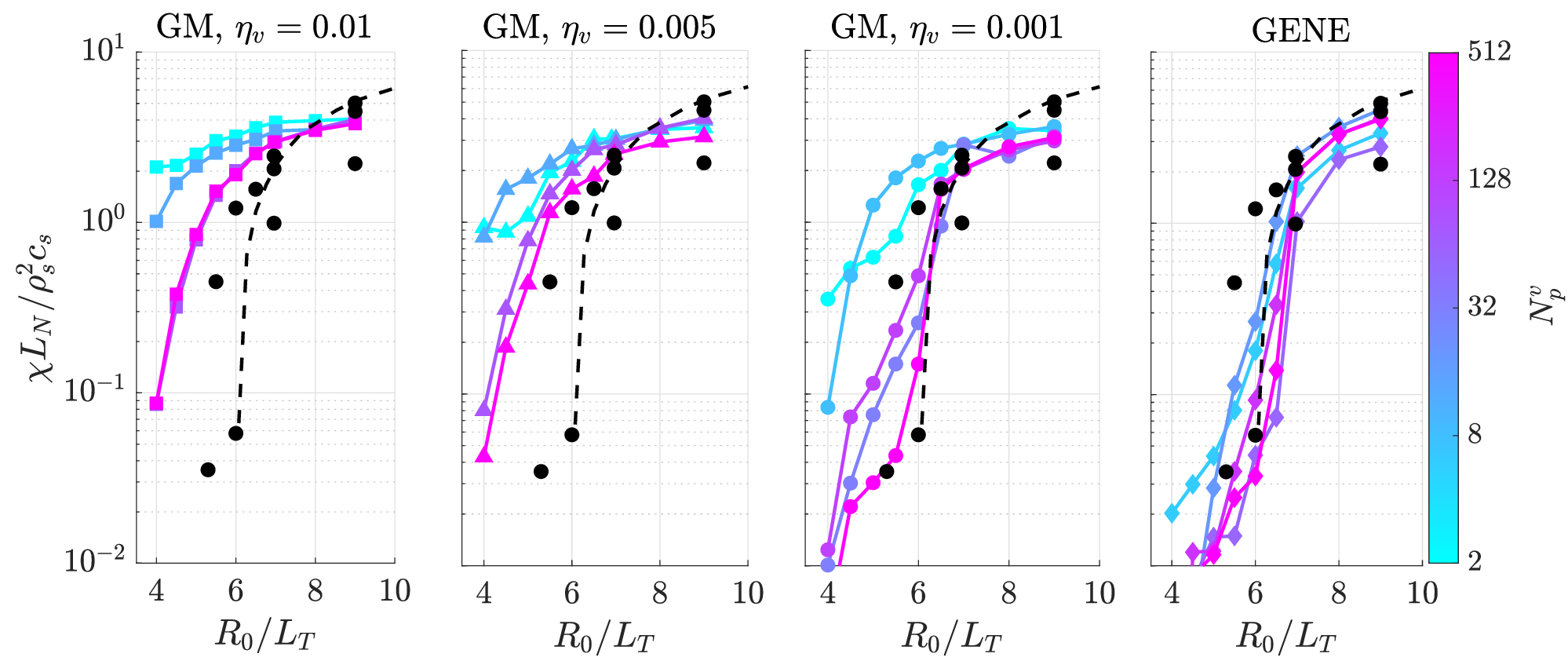}
    \caption{Heat diffusivity as a function of the temperature gradient strength $R/L_T$ obtained with the GM approach varying the intensity of the numerical velocity dissipation, $\eta_{v}=0.01$, $\eta_{v}=0.005$ and $\eta_{v}=0.001$ and obtained with GENE. 
    The colors indicate the number of points in the velocity space, $N_p^v=(P+1)\times(J+1)$ with $(P,J)=(2,1)$, $(4,2)$, $(8,4)$, $(16,8)$ and $(30,15)$ for the GM approach and $N_p^v=N_{v\parallel}\times N_\mu=8\times4$, $16\times8$, $32\times16$, $42\times24$ for GENE.
    The results of GK and PIC simulations in \cite{Dimits2000ComparisonsSimulations} are reported in black.}
    \label{fig:42_Dimits_convergence}
\end{figure}
Finally, we investigate the effect of numerical dissipation in velocity space with a convergence study focused on the heat diffusivity, carried out using three different dissipation intensities: $\eta_{v}=0.05$ (Fig. \ref{fig:42_Dimits_convergence}a), $\eta_{v}=0.01$ (Fig. \ref{fig:42_Dimits_convergence}b) and $\eta_{v}=0.001$ (Fig. \ref{fig:42_Dimits_convergence}c). 
We compare the results with those obtained using GENE (Figure \ref{fig:42_Dimits_convergence}d) as well as the results reported in \cite{Dimits2000ComparisonsSimulations} for PIC and GK codes.
Figure \ref{fig:42_Dimits_convergence} shows that the lowest GM sets, $(P,J)=(2,1)$ and $(4,2)$, significantly overestimates the transport values compared \cite{Dimits2000ComparisonsSimulations} and GENE, for all dissipation levels and particularly for the lowest background gradient considered. 
This can be attributed to the overestimate of the linear growth rate, observed in particular at the lowest gradient levels. 
On the other hand, GENE linear results do not exhibit this overestimate at low resolution, which explains that GENE results do not overestimate the transport at low resolution.

For the $(8,4)$ and larger GM sets, a good agreement with the results obtained by the various GK codes presented in \cite{Dimits2000ComparisonsSimulations} and GENE is observed only for $\eta_{v}=0.001$.
\reviewi{In scenarios with low velocity dissipation and close to marginal stability, the GM approach does not outperform GENE in the low resolution limit. 
This implies the existence of non-Maxwellian velocity space structures in the perturbed distribution function, which are challenging to resolve with only a few Hermite-Laguerre modes. 
This aligns with our previous study on the Dimits shift in a Z-pinch, where the velocity distributions are compare in more details \citep{Hoffmann2023GyrokineticOperators}.}
At higher dissipation levels, the GM approach converges faster but yields inaccurate values, in particular when the gradient level becomes smaller than in the CBC.
It is worth noting that the nonlinear GM results converge faster as the gradient is increased, similarly to the linear growth rate.

\section{Collisional effects on the Dimits shift}
\label{sec:coll_dimits_shift}
We explore the impact of collisions on the Dimits shift using advanced GK linearized collision operators.
We compare the effect of the linear GK Dougherty (DGGK), Sugama (SGGK) and Landau (LDGK) collision operators on the transport level.
We first consider the same parameters as the CBC and set the temperature background level to $\kappa_T=5.3$. 
We vary the collision frequency parameter around $\nu \sim 0.005$, which corresponds to experimental value of the DIII-D discharge considered for the CBC \citep{Lin1999EffectsTransport}. 
We then consider the impact of collisions for different values of $\kappa_T$\modifii{, setting $\eta_{v}=0$}.

Motivated by the observed correlation between the level of heat transport and the ITG linear growth rate observed in the collisionless case (see Sec. \ref{sec:coll_dimits_shift}), we first study the impact of our different collision models on the linear growth rates.
\begin{figure}
    \centering    \includegraphics[width=1.0\linewidth]{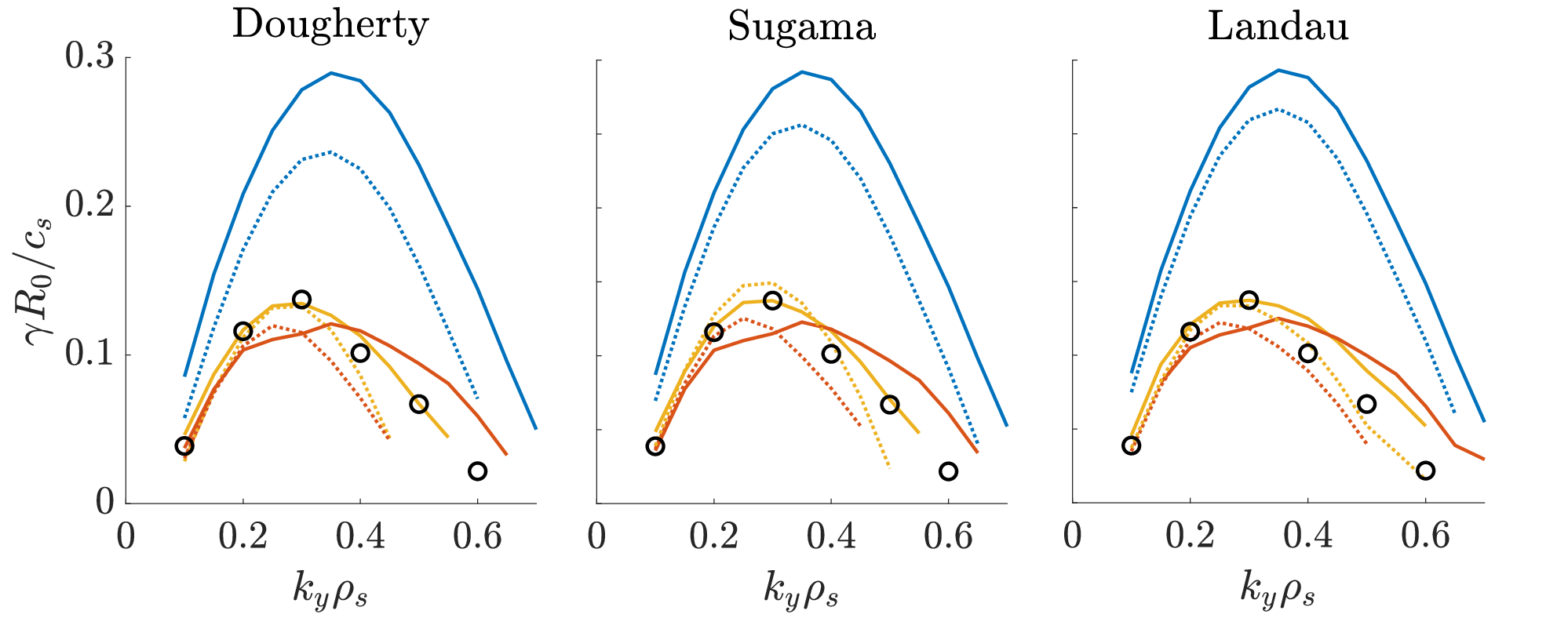}
    \caption{Collisional study of the ITG growth rate at $\kappa_T=5.3$ with the $(P,J)=(4,2)$ (blue), $(P,J)=(8,4)$ (orange) and $(P,J)=(16,8)$ (green) basis. 
    Different collision operators are compared with collision frequencies $\nu=0.05$ (dashed) and $\nu=0.005$ (solid). 
    The converged collisionless result are also shown (black circles).}
    \label{fig:51_lin_comparison}
\end{figure}

These are presented in Fig. \ref{fig:51_lin_comparison}, with the collision frequency ranging from $\nu=0.05$ to $\nu=0.005$, and the polynomial basis varying from $(P,J)=(4,2)$ to $(16,8)$. 
In agreement with the collisionless case, we observe that the $(4,2)$ basis tends to overestimate the growth rate for both collision frequencies across all three collision models. 
However, faster convergence is observed compared to the collisionless scenario\reviewi{, which is a consequence of the fact that the GK model tends towards a fluid limit at high collisionality}. 
Specifically, the growth rates obtained with the $(P,J)=(8,4)$ basis closely approach those obtained with the $(16,8)$ basis.
This is not the case in the collisionless regime. 
Notably, we find that the growth rates do not exhibit significant variations among the different collision operators and are weakly sensitive to the collision frequency for large GM sets.
%
\begin{figure}
    \centering
    \includegraphics[width=0.49\linewidth]{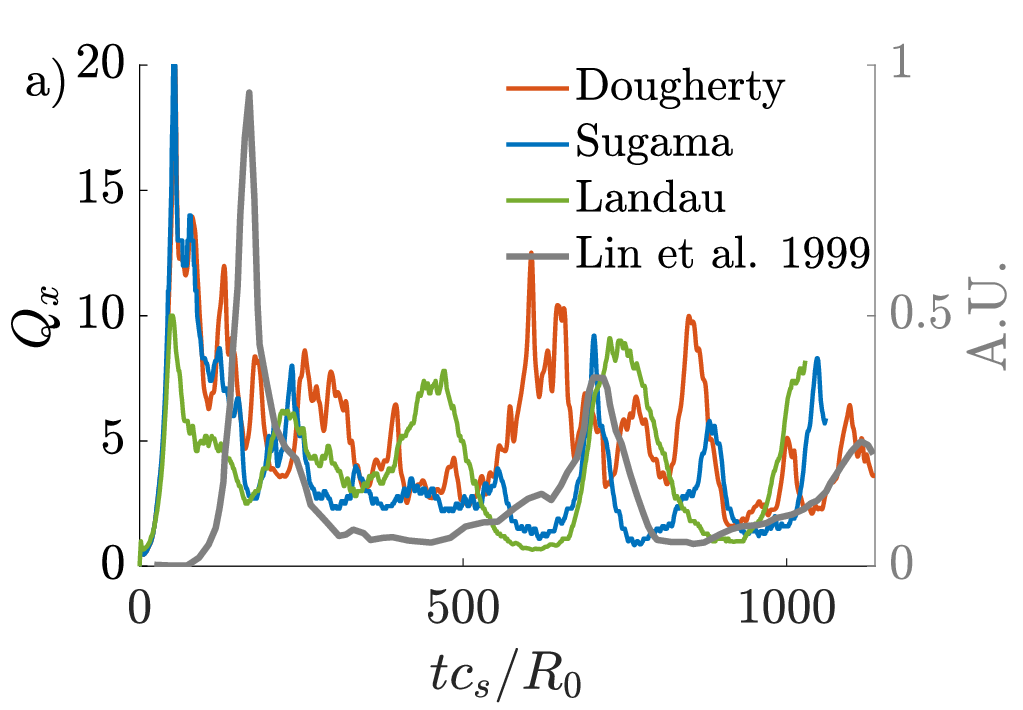}
    \includegraphics[width=0.49\linewidth]{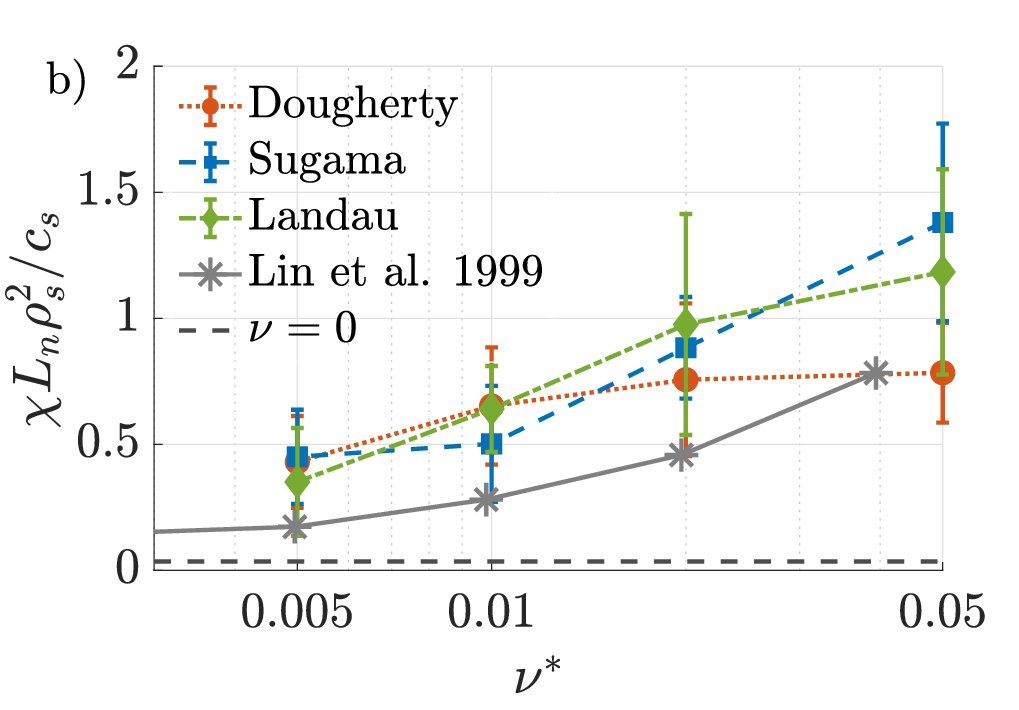}
    \caption{Study of collision effects on the heat flux with the GM approach and the Dougherty (orange), Sugama (blue) and Landau (green) operators for $\kappa_T=5.3$ for $\nu=0.005$. 
    The heat flux time traces (a) and the average heat diffusivity (b) are compared with the results from \cite{Lin1999EffectsTransport}, reported in gray.
    The value of the heat diffusivity in the collisionless case is also shown (black dashed line).
    The error bars represent the standard deviation of the heat flux.}
    \label{fig:52_nlin_nu_scan}
\end{figure}
We now perform nonlinear simulations with the GM code at $\kappa_T=5.3$ for a set of collision frequencies, namely $0.005\leq\nu\leq 0.05$ with the $(P,J)=(16,8)$ polynomial basis.
In Fig. \ref{fig:52_nlin_nu_scan}a, we present the heat flux time traces obtained by \cite{Lin1999EffectsTransport}, and the GM approach using the three collision operators considered here. 
Remarkably, we observe similar results across all collision models, with bursts increasing the transport by a factor five with respect to the baseline level, as also observed by \cite{Lin1999EffectsTransport}.
\reviewi{It is worth mentioning that these bursts are also observed in GK simulations \citep{Kobayashi2015,Peeters2016Gradient-drivenThreshold,Hallenbert2022} as well as reduced turbulence modelling \citep{Qi2020,Ivanov2020ZonallyTurbulence,Ivanov2022DimitsTurbulence}}.
Figure \ref{fig:52_nlin_nu_scan}b shows the average heat diffusivity as a function of collision frequency for the different collision models, compared with the results from \cite{Lin1999EffectsTransport}. 
We note a good agreement in terms of trend and transport amplitude, particularly considering that the results from \cite{Lin1999EffectsTransport} are based on global PIC simulations performed with the GTC code \citep{Lin1998TurbulentSimulations}.

Our results show that the choice of collision model does not appear to significantly affect the heat flux at collision frequencies close to the physical values observed in the core of the DIII-D tokamak. 
The choice of collision models impacts the transport only when the collision frequency is ten times higher than the physically relevant value.
This contrasts with \cite{Hoffmann2023GyrokineticOperators}, where a noticeable effect of the choice of the collision operator is observed also at low collision frequency.
We attribute this difference to two main factors. 
First, the ITG instability exhibits a stability spectrum with predominantly stable small-scale wavelengths, thus considerably less affected by collisions, with respect to the entropy mode considered by \cite{Hoffmann2023GyrokineticOperators} that develops also on short scales.
Second, the collision operators used here account only for ion-ion collisions while electron-ion collisions are considered in the entropy mode investigations.

\begin{figure}
    \centering
    \includegraphics[width=1.0\linewidth]{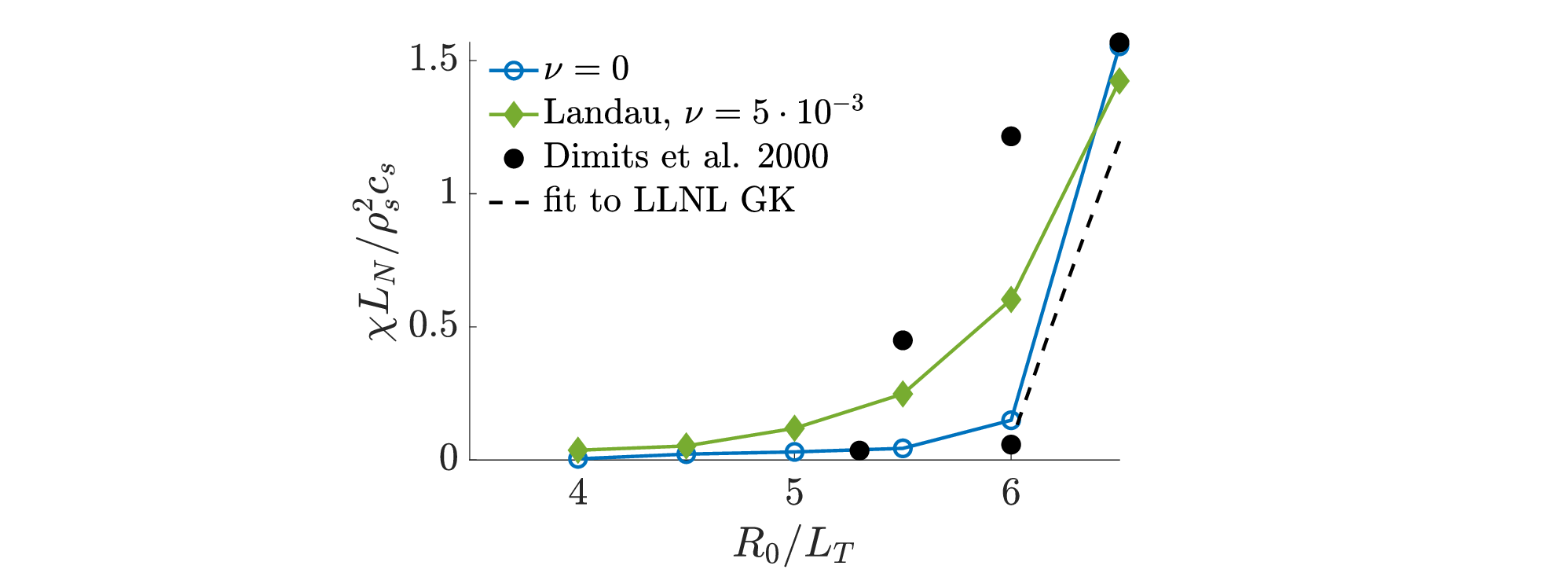}
    \caption{Heat diffusivity as a function of the temperature gradient strength $R/L_T$ obtained with the GM approach in the collisionless limit (blue circles) and for $\nu=0.005$ with the Landau GK collision operator (green diamonds).
    \reviewi{We report the collisionless results (black dots) of \cite{Dimits2000ComparisonsSimulations} and the fit used therein to obtain the threshold value $\kappa_T=6$ (dashed line).}}
    \label{fig:53_collisional_dimits_shift}
\end{figure}
Finally, we study the effect of the Landau collision operator on the Dimits shift at $\nu_{LDGK}=0.005$.
As shown on Fig. \ref{fig:53_collisional_dimits_shift}, the collisionality tends to smooth out the transition between fully developed turbulence ($\kappa_T\gtrsim 6$) and zonal flows dominated saturated states ($\kappa_T\lesssim 6$), where collisions have a stronger effect on the transport level.
\reviewiii{This observation is also made in \cite{Peeters2016Gradient-drivenThreshold}, where numerical dissipative terms are used to obtain the smoothing of the Dimits shift.
This suggests that advanced collision operators are not required to accurately resolve the CBC with an adiabatic electron model, as zonal flow are mostly damped through diffusion in the configuration and velocity spaces.}

\section{Conclusions}
\label{sec:conclusion}
In the present study, we report on a benchmark and convergence analysis of the GM approach in the local flux-tube $\delta f$ framework, with a specific focus on the CBC and the Dimits shift. 
\modifii{The GM approach converges faster than the GENE code in yielding the correct linear growth rate of the low $k_y\rho_s$ modes, and therefore a mixing-length estimate of the turbulent transport.}
In the nonlinear case, our findings highlight that the GM approach accurately captures the nonlinear dynamics of the CBC, while using significantly fewer velocity space points compared to the GENE code. 
We observe that increasing the intensity of velocity dissipation improves the convergence rate, albeit with a \reviewii{$30\%$} discrepancy in the saturated heat flux value.
In addition, the GM approach successfully replicates the GK results reported in \cite{Dimits2000ComparisonsSimulations}, in contrast to the GF models. 
The Dimits shift is obtained accurately with a comparable number of moments as the number of GENE velocity grid points, observing slower convergence as the system approaches marginal stability. 
Nonetheless, it is important to highlight that the GM approach effectively captures the width of the Dimits shift also at low velocity space resolution. 
This highlights that the primary limitation of the GM approach lies in the convergence of the linear stability threshold rather than in the nonlinear dynamics.
Thus, evolving additional GMs in the CBC acts mostly as a fine tuning of the linear ITG instability.
\reviewi{It also confirms that a simple model based on the $E\times B$ advection, as in \cite{Ivanov2022DimitsTurbulence}, is sufficient to predict correctly the dynamics in the Dimits region.
Consequently, the balance between Reynolds and diamagnetic stresses, present in the  $E\times B$ advected temperature assumption \citep{Sarazin2021KeyPlasmas}, is maintained when increasing the number of evolved GMs in the CBC.
However, one must recall that this affirmation is challenged by GK simulations based on the global code GYSELA \citep{Sarazin2021KeyPlasmas}, suggesting that the disagreement between GYSELA and \cite{Ivanov2022DimitsTurbulence} simulations resides in features of the models such as the local assumption or the boundary conditions.
}

In the collisional case, our analysis of transport within the Dimits window ($\kappa_T=5.3$) reveals a minimal influence of collision on the ITG growth rate around a physical collision frequency of $\nu \sim 0.005$. 
This observation holds for all the three GK linear collision models examined, namely Dougherty, Sugama, and Landau. 
While the GM approach exhibits good agreement with the global PIC results reported in \cite{Lin1999EffectsTransport}, when varying the collision frequency, the collision model itself does not exhibit a significant impact around the DIII-D core relevant collision frequency.
We attribute this result to the long-wavelength nature of the ITG instability and to the adiabatic electron model, which prevents the impact of electron-ion collisions on the plasma dynamics.

In a broader context, the present study is a stepping stone towards the efficient modelling of plasma turbulence in the edge region.
In fact, as confirmed by the present work, the high collisionality, combined with large background gradients typical of the edge, are ideal conditions to apply the GM approach.

\section*{Acknowledgements}
The authors acknowledge helpful discussions with J. Ball, S. Brunner, A. Vol\v{c}okas, N. Mandell and P. Giroud-Garampon.
The simulations presented herein were carried out in part on the CINECA Marconi supercomputer under the TSVVT422 project and in part at CSCS (Swiss National Supercomputing Center).
This work has been carried out within the framework of the EUROfusion Consortium, via the Euratom Research and Training Programme (Grant Agreement No 101052200 — EUROfusion) and funded by the Swiss State Secretariat for Education, Research and Innovation (SERI). Views and opinions expressed are however those of the author(s) only and do not necessarily reflect those of the European Union, the European Commission, or SERI. Neither the European Union nor the European Commission nor SERI can be held responsible for them.

\section*{Declaration of interests}
The authors report no conflict of interest.

\newpage
\appendix

\bibliographystyle{jpp}
\bibliography{references}
\end{document}